\documentclass[preprint,10pt]{elsarticle}

\usepackage[a4paper, left=0.75in, right=0.75in, top=1in, bottom=1in]{geometry}
\usepackage{amsmath}
\makeatletter
\def\maketag@@@#1{\hbox{\m@th\normalfont\normalsize#1}}
\makeatother
\usepackage{amssymb}
\usepackage{graphicx}
\usepackage{subcaption}
\usepackage{mathtools}
\usepackage{indentfirst}
\usepackage{xcolor}
\usepackage{hyperref}
\hypersetup{
	bookmarks=true,
	bookmarksnumbered=true,     
	bookmarksopen=true,         
	bookmarksopenlevel=1,       
	hidelinks,
	pdfstartview=Fit,           
	pdfpagemode=UseOutlines,    
}

\numberwithin{equation}{section}
\numberwithin{table}{section}
\numberwithin{figure}{section}

\newcommand\numberthis{\addtocounter{equation}{1}\tag{\theequation}}
\let\vec\mathbf

\makeatletter
\def\ps@pprintTitle{%
	\let\@oddhead\@empty
	\let\@evenhead\@empty
	\def\@oddfoot{}%
	\let\@evenfoot\@oddfoot}
\makeatother

\begin{document}

\begin{frontmatter}



\title{Multiple scales analysis of out-of-plane and in-plane vibrations of a wind turbine blade}

\author[UW]{Yash Talwekar\corref{cor}}
\ead{yasht16@uw.edu}

\author[Birla]{Aswan Abdul Razak}
\author[Birla]{Amol Marathe}
\ead{amolmm@pilani.bits-pilani.ac.in}

\address[UW]{Department of Mechanical Engineering, University of Washington, Seattle, WA 98105, USA}
\address[Birla]{Department of Mechanical Engineering, Birla Institute of Technology and Science, Pilani, Rajasthan 333031, India}

\cortext[cor]{Corresponding Author}
\begin{abstract}
		Steady-state response of an isolated horizontal axis wind turbine blade undergoing out-of-plane and in-plane vibrations are studied considering the two cases separately. Equations of motion for both cases are sought by modeling the blade as cantilevered Euler-Bernoulli beam and applying the Lagrangian formulation. Taking into account the effects of blade rotation on gravitational and aerodynamic forces and allowing the blade to undergo large deformations, various nonlinearities arise in the system resulting in superharmonic resonances. Taking practically relevant numerical values of parameters, we suitably order the parameters for their smallness. We then obtain the frequency responses corresponding to primary and superharmonic resonances by applying the method of multiple scales up to fourth and third orders for out-of-plane and in-plane cases respectively. Comparisons are drawn with the results obtained using the method of harmonic balance.
\end{abstract}

\begin{keyword}
Wind Turbine Blade
\sep
Frequency Response
\sep
Superharmonic Resonances
\sep
Multiple Scales
\sep
Harmonic Balance
\end{keyword}

\end{frontmatter}


\section{Introduction}
The demand for efficient energy extraction from renewable sources such as wind is increasing steadily. This technically requires wind turbines with larger diameters and higher rotating speeds. This necessitates to take into account the rotor models with longer blades that undergo large deformations. Subsequently nonlinear effects become prominent and have to be incorporated in the model. Although numerical studies based on a full order nonlinear model (e.g. a model based on finite elements) offers a detailed account of the rotor blade dynamics, such studies typically become very expensive computationally. Analysis based on a reduced order model (e.g. one d.o.f) is a good starting point to gain qualitative insight into the interaction of various physical factors and studying the overall system performance.

There have been various numerical as well as analytical studies concerning the vibration characteristics of a rotor blade modeled using a single d.o.f. Most of the studies pertaining to nonlinear analysis of such a model include variations in aerodynamic and gravitational loads due to the blade's rotation. Modeling the blade as a rigid body with three rotational degrees of freedom about the hinges at its root considering quasi-static aerodynamics, Chopra et al.\cite{IChopra} found the frequency response using the harmonic balance method and carried out self-excited flutter analysis. If we model the blade to be flexible, curvature-induced nonlinearities arise in the model. Inoue et al.\cite{Inoue} investigated the occurrence and characteristics of superharmonic resonances in a wind turbine blade through numerical simulation and theoretical analysis, and confirmed the same experimentally. Ramakrishnan et al.\cite{Feeny1, Feeny2} considered effects of cyclic transverse loading due to wind shear, tower shadowing and cyclic gravitational axial loading in case of in-plane vibrations to obtain equation of motion with both external and parametric excitations. With this motivation, they applied the method of multiple scales to a forced nonlinear Mathieu equation. Larsen et al.\cite{Larsen} studied the dynamic behaviour of a blade subjected to harmonically varying support point motion, which introduces additional nonlinearities in the form of inertial loading whilst considering the coupling between fundamental and edgewise vibrations. Li et al.\cite{Li} derived mathematical model for a rotor blade undergoing flapwise, edgewise, torsional and axial vibrations including various factors such as coning, twist and hub offset using aerodynamic loading obtained via unsteady airfoil theory. Many studies also focus on the control and mitigation of these vibrations such as that by Staino et al.\cite{Staino}, where they consider use of actuators which generate controllable forces on the blade while formulating the model.

In this paper, we study the steady-state response of an isolated elastic blade of a horizontal axis wind turbine undergoing out-of-plane (flapwise) as well as in-plane (edgewise) vibrations. We do not consider a coupled system, but rather study the two scenarios separately. A horizontal axis wind turbine, as the name suggests, has rotor blades in the vertical plane rotating about turbine's main shaft. Due to this motion there is a periodic influence of gravity on the blades. Wind velocity varies with respect to height from the ground surface (where it is zero) due to the wind shear. This results in the periodic aerodynamic forces acting on the blades. Such variation of loading of the turbine blade with respect to its angular position makes the analysis different from that of a rotor system with blades rotating in the horizontal plane, such as a helicopter rotor.
In both cases (in- and out-of-plane), the blade is approximated by a cantilever beam undergoing large deformations expressed in terms of the fundamental bending mode only. We follow Inoue et al.\cite{Inoue} in deriving the equations of motion for the reduced order model using the Lagrangian formulation. Due to rotation, gravity acts in the form of parametric excitation in both cases, and also in the form of external excitation in case of the in-plane vibrations. Considering wind shear, aerodynamic loading acts as external periodic excitation. We consider the expression for aerodynamic loading in the edgewise direction from Li et al.\cite{Li}. Due to various nonlinearities present in the system, superharmonic resonances are observed. We use the method of harmonic balance in combination with arc-length based continuation scheme to obtain the frequency response near different resonances. We then identify a small parameter in the equations of motion to apply the method of multiple scales. We carry out the multiple scales analysis up to fourth order and third order in case of out-of-plane and in-plane vibrations respectively. Considering each resonance separately, we obtain the corresponding slow flow. Then using the numerical continuation scheme, we get the frequency response. The same is compared with that obtained using harmonic balance.
\section{Out-of-plane vibrations}
\subsection{System Description}
\begin{figure}[h]
	\centering
	\begin{subfigure}[b]{0.47\textwidth}
		\includegraphics[width=\textwidth]{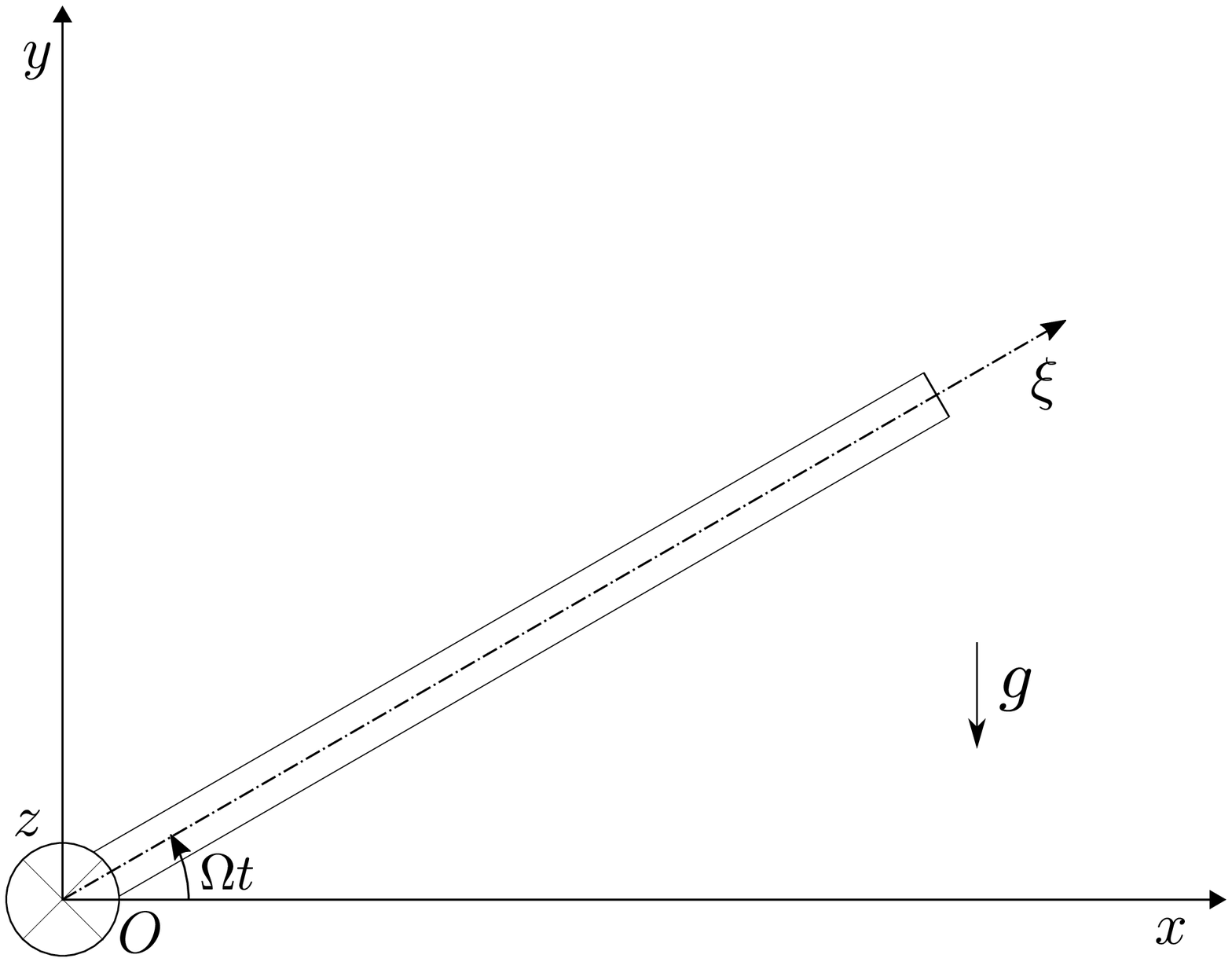}
		\caption{}
		\label{fig:Wtmodel2a}
	\end{subfigure}
	\begin{subfigure}[b]{0.3\textwidth}
		\includegraphics[width=\textwidth]{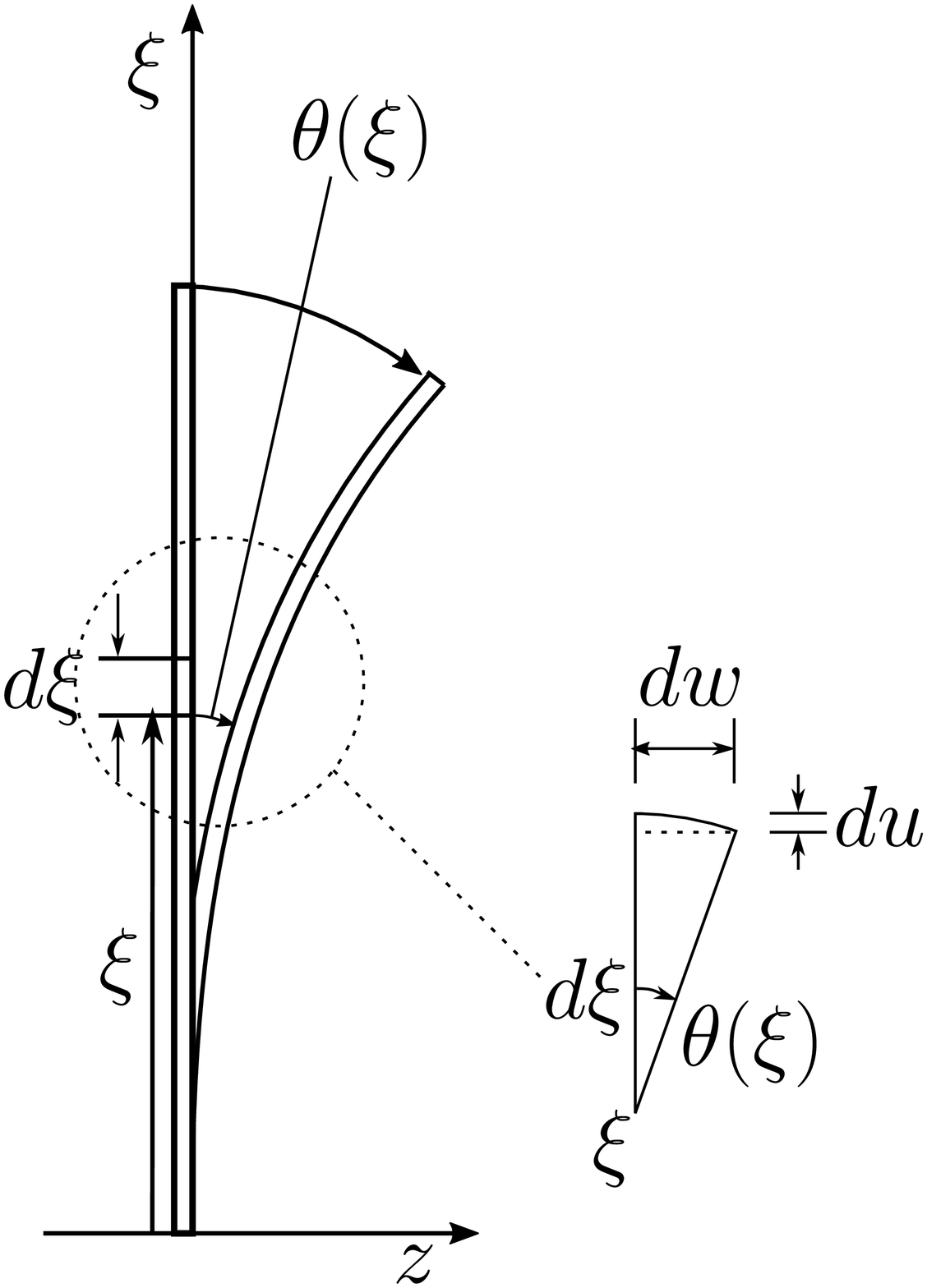}
		\caption{}
		\label{fig:Wtmodel2b}
	\end{subfigure}
	\caption[Coordinate systems]{(a) Coordinate systems, (b) Displacement in the out of plane direction.}
\end{figure}

The model of the horizontal axis wind turbine blade and the corresponding equation of motion under consideration refer to the work of Inoue et al.\cite{Inoue}.
The blade is modelled as an Euler-Bernoulli beam with one end fixed to the horizontal main shaft. Only the vibrations of first bending mode in the out-of-plane direction are considered and the same due to higher modes as well as in-plane and torsional modes are ignored. The wind force acting on the rotating blade in the out-of-plane direction varies with height. It is assumed that the axis of blade's rotation is aligned with the wind direction. The origin of the inertial frame of reference $O$-$xyz$ is chosen at blade's installation position as can be seen from  Fig.\ref{fig:Wtmodel2a}. Let $\Omega$ be the constant angular velocity of the blade about $z$-axis and let $\xi$-axis aligned with $x$-axis at $t=0$ be rigidly attached to the blade.

\subsubsection{Lagrangian and equation of motion}
Approximating the blade as a uniform beam of length $l$ and the mode-shape corresponding to the first mode of the rotating beam by that of the non-rotating beam, i.e, uniform cantilevered Euler-Bernoulli beam, we assume \cite{Inoue}

\begin{align} \label{eq:hxi}
\mathllap{\psi(\xi)} = \big(\sin(\lambda_{1}) + \sinh(\lambda_{1})\big)\Big(\cos(\lambda_1 \frac{\xi}{l}) - \cosh(\lambda_1 \frac{\xi}{l})\Big)
- \big(\cos(\lambda_1) + \cosh(\lambda_1)\big)\Big(\sin(\lambda_1 \frac{\xi}{l}) - \sinh(\lambda_1 \frac{\xi}{l}) \Big),
\end{align}
where $\xi$ denotes position along $\xi$-axis and
$\lambda_{1} = 1.875\ldots\;$ is the eigenvalue corresponding to the first bending mode of vibration\cite{Meirovitch},\cite{Douglas}.
Consider a point along the beam located at $\xi.$ Its co-ordinates at $t=0$ are $(\xi, 0 , 0).$ Let $(x_\xi, y_\xi, z_\xi)$ denote the co-ordinates at time t. Deflection of this point is represented as
\begin{equation*}
w(\xi, t) = \frac{\psi(\xi)}{\psi(l)} z_l(t) ,
\end{equation*}
where $z_l(t)$ is the blade-tip displacement in the $z$-direction.
With the blade deflected by an angle $\theta(\xi) = {\partial w(\xi,t)}/{\partial \xi}$ (Fig.\ref{fig:Wtmodel2b}), displacement of this point in longitudinal direction is
\begin{align} \label{eq:uxi}
u(\xi,t)=\alpha(\xi) z_l(t)^2
\quad 
\mbox{with}
\quad  
\alpha(\xi)=\frac{1}{2\psi(l)^2}\int_{0}^{\xi}\left(\frac{d\psi}{d\xi}\right)^2 d\xi.
\end{align}
When the blade is at an angle $\Omega t,$ co-ordinates of this point are given by
\small
\begin{align} \label{eq:coords}
x_\xi=\big(\xi-u(\xi)\big)\cos(\Omega t) =\big(\xi-\alpha z_l^2\big)\cos(\Omega t) ,
\quad
y_\xi=\big(\xi-u(\xi)\big)\sin(\Omega t) =\big(\xi-\alpha z_l^2\big)\sin(\Omega t) , 
\quad
z_\xi=\frac{\psi(\xi)}{\psi(l)}z_l .
\end{align}
\normalsize
Potential energy of the blade is not only due to its deflection but due to its vertical position as well. Kinetic and potential energies of the blade are then defined as
\begin{align} \label{eq:TUexp}
T &=\frac{1}{2}\rho A \int_{0}^{l}(\dot{x}_\xi^2 +\dot{y}_\xi^2+\dot{z}_\xi^2) d\xi 
\qquad \mbox{and} \qquad
U=\frac{1}{2}\int_{0}^{l}EI\left(\frac{\partial^2{w}}{\partial\xi^2}\right)^2d\xi+\rho Ag \int_{0}^{l}y_\xi d\xi ,
\end{align}
where $\rho,A,E$ and $I$ are mass density, cross sectional area, Young's modulus and area moment of inertia of the blade respectively. And $g$ denotes the acceleration due to gravity.
Substituting Eq.\eqref{eq:coords} in Eq.\eqref{eq:TUexp}, we obtain kinetic energy as
\begin{gather*}
T =\frac{1}{2}\rho A\left(\alpha_1 z_l^2\left(\Omega^2 z_l^2 + 4\dot{z}_l^2\right)+ \frac{l^3 \Omega^2}{3}-2\alpha_2\Omega^2 z_l^2+\alpha_3\dot{z}_l^2\right) ,
\\
\mbox{with} \quad
\alpha_1=\int_{0}^{l}\alpha^2 d\xi=0.0718,\;
\alpha_2=\int_{0}^{l}\xi \alpha d\xi=0.1491 \quad \mbox{and} \quad
\alpha_3=\frac{1}{\psi^2(l)}\int_{0}^{l}\psi^2d\xi=0.2499 .
\end{gather*}
Similarly potential energy is obtained as
\begin{gather*}
U = \frac{1}{2}EI\beta_4 z_l^2+ g\rho A\left(\frac{l^2}{2}-\beta_5 z_l^2\right) \sin(\Omega t),
\\
\mbox{with} \quad
\beta_4=\frac{1}{\psi(l)^2} \int_{0}^{l}\left(\frac{\partial^2{\psi}}{\partial\xi^2}\right)^2d\xi =3.0905,
\qquad
\mbox{and}
\qquad
\beta_5 =\int_{0}^{l}\alpha d\xi =0.1963 .
\end{gather*}

Non-conservative forces acting on the blade are due to viscous damping and the external force due to wind. We assume Rayleigh's dissipation function for the system to be
\begin{equation*}
D= \frac{1}{2}c_l\dot{z}^2_l =\int_{0}^{l}\frac{1}{2}c_\xi\dot{z}^2_\xi d\xi,
\end{equation*}
where $c_\xi$ is the damping coefficient per unit length and $c_l =  c_\xi \alpha_{3}$ is the total damping coefficient. Assuming the effect of wind shear and the blade's instantaneous angular position as harmonic, we express the aerodynamic force as \cite{Inoue}
\begin{equation*}
Q_l=Q_c+\Delta Q\sin(\Omega t),
\end{equation*}
where $Q_c$ is the wind force at height corresponding to the main shaft of the wind turbine and $\Delta Q$ is the maximum change in wind force when the blade is at the top position.

\begin{figure}[h]
	\centering
	\includegraphics[width=0.6\textwidth]{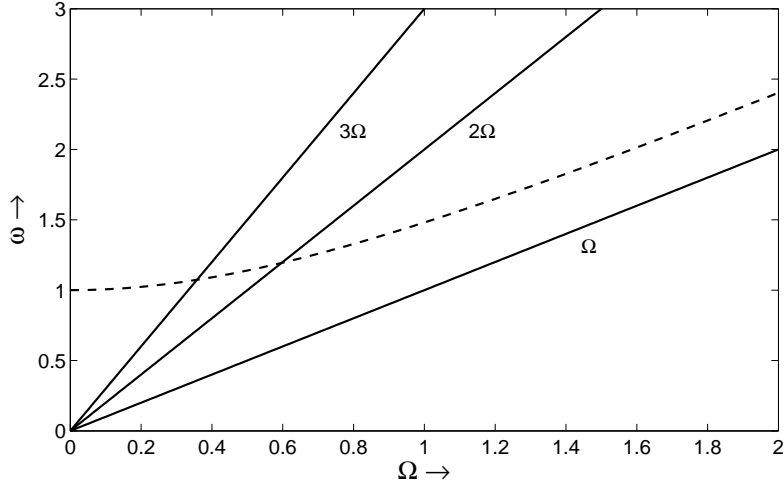}
	\caption{Natural frequency diagram}
	\label{fig:znatfreq}
\end{figure}

For a generalised co-ordinate $q,$ Euler-Lagrange equation of motion is written as
\begin{equation} \label{eq:Lagrange}
\dfrac{d}{dt}\bigg(\frac{\partial L}{\partial \dot{q}}\bigg) - \frac{\partial L}{\partial q} + \frac{\partial D}{\partial \dot{q}} = Q_{ext} ,
\end{equation}
where, $L=T-U$ is the Lagrangian and $Q_{ext}$ is forcing external to the system. Taking $z_l$ as the generalized co-ordinate and $Q_{ext} = Q_l$ in Eq.\eqref{eq:Lagrange}, we get the dimensional equation of motion of an isolated elastic blade as
\begin{equation}
\label{WTeq}
\rho A (4\alpha_1z^2_l+\alpha_3) \ddot{z}_l+c_l\dot{z}_l+\big(4\alpha_1\rho A \dot{z}_l^2+2\rho A \Omega^2(\alpha_2-\alpha_1z^2_l) 
+EI\beta_4-2g\rho \beta_5A \sin(\Omega t)\big)z_l=Q_c+\Delta Q \sin(\Omega t) .
\end{equation} 
Dimensions and structural properties of the blade are taken from \cite{Inoue} and are given in Table \ref{table1}. Parametric excitation present in Eq.\eqref{WTeq} is due to gravity and is primarily responsible for large amplitude peaks in superharmonic resonances. Blade rotation resulting in centrifugal stiffening is taken into account by the term $2\rho\alpha_{2}A\Omega^2 z_l.$ Geometric nonlinearities are taken care of in the form of cubic nonlinear terms in the equation. Referring to Inoue et al.\cite{Inoue}, solution for the free vibrations of the blade with static deflection is assumed as
\begin{equation*}
z_l(t) = z_{l0} + z_{l1} \sin(\omega t),
\end{equation*}
where $z_{l0}$ is the static deflection, $z_{l1}$ is the free vibration amplitude and $\omega$ is the natural frequency.
Substituting the above in Eq.\eqref{WTeq} with the condition $c_l=0$, $Q_c=\Delta Q = 0$ and ignoring the parametric excitation, expanding and combining terms, and equating the coefficient of $\sin(\omega t)$ to zero gives us the expression for the natural frequency as
\begin{equation}
\label{eq:natfreq}
\omega = \sqrt{\frac{2 E I \beta_4 + \rho A \Omega^2\big(4\alpha_2 - 3\alpha_1(4 z_{l0}^2 + z_{l1})\big)}{2 \rho A (\alpha_3 + 2\alpha_1( 2 z_{l0}^2 + z_{l1}^2))}}.
\end{equation}
\begin{table}[h]
	\parbox[b]{.45\linewidth}{
		\centering
		\begin{tabular}[b]{|l|r|}
			\hline \textbf{Parameter} & \textbf{Value} \\ 
			\hline Blade length, $l$ & $1.0  m$ \\ 
			\hline Blade thickness, $d$ & $0.0025   m$ \\ 
			\hline Blade width, $b$ & $0.04   m$ \\ 
			\hline Young's modulus, $E$ & $200   GPa$ \\ 
			\hline Mass Density, $\rho$ & $7870   kg/m^3$ \\ 
			\hline Damping coefficient, $c_l$ & $0.001   Ns/m$ \\ \hline $Q_c$ & $0.3   N$ \\ 
			\hline $\Delta Q$ & $0.1   N$ \\ 
			\hline 
		\end{tabular} 
		\caption{Parameter values of the elastic blade.}
		\label{table1}
	}
	\hfill
	\parbox[b]{.45\linewidth}{
		\centering
		\begin{tabular}[b]{|c|c|}
			\hline
			\textbf{Parameter} & \textbf{Value} \\
			\hline
			${\alpha}_1$ & $7.1824 \times 10^{-6}$\\
			\hline ${\alpha}_2$ & $1.1934$\\
			\hline ${c}$ & $3.97 \times 10^{-4}$		\\
			\hline ${g}$ & $0.094239$		\\
			\hline ${Q}_c$	& $3.73$	\\
			\hline $ \Delta {Q}$ & $1.24$ \\
			\hline
		\end{tabular}
		\caption{Dimensionless parameters in Eq.\eqref{eq:mroe}.}
		\label{table:dimpar1}	
	}
\end{table}
We divide Eq.\eqref{WTeq} by $\rho \alpha_3 d \omega_0^2 A ,$ where $\omega_0=\sqrt{E I \beta_4 /\alpha_3\rho A}$ is the natural frequency in the non-rotating and small displacement condition and $d$ is thickness of the blade. Introducing dimensionless versions of the variables and parameters such as
\begin{gather*} 
\bar{t}=\omega_0t,\; 
x=\frac{z_l}{d} \bigg(x_0=\frac{z_{l0}}{d}, \; x_1=\frac{z_{l1}}{d}\bigg),\; 
\bar{\alpha}_1=\frac{4 \alpha_1 d^2}{\alpha_3}, \;\bar{\alpha}_2=\frac{2 \alpha_2}{\alpha_3},\; \bar{\Omega}=\frac{\Omega}{\omega_0},\\
\bar{c}=\frac{c_l}{\sqrt{\alpha_3 \rho A EI\beta_4 }},\; \bar{g}=\frac{2 g \rho A \beta_5}{EI \beta_4},\; \bar{Q}_c=\frac{Q_c}{d EI \beta_4},\; \Delta\bar{Q}=\frac{{\Delta}Q}{d EI \beta_4},
\numberthis \label{eq:dimpar}
\end{gather*}
we obtain Eq.\eqref{WTeq} in dimensionless form as
\begin{equation} \label{eq:mroe}
\big( 1+\bar{\alpha}_1 x^2 \big)\ddot{x} + \bar{c}\dot{x}
+  \Big(\bar{\alpha}_1\dot{x}^2
+ \bar{\Omega}^2 \big( \bar{\alpha}_2-\frac{1}{2} \bar{\alpha}_1 x^2\big) 
+1 - \bar{g}\sin \big( \bar{\Omega} \bar{t} \big) 
\Big) x 
= \bar{Q}_c + \Delta \bar{Q} \sin \big( \bar{\Omega} \bar{t} \big) ,
\end{equation}
where overdot represents differentiation with respect to the non-dimensional time $\bar{t}.$ Hereafter, we drop overbar notation for all nondimensional quantities. Numerical values of the parameters in Eq.\eqref{eq:mroe} are given in Table \ref{table:dimpar1}. Dimensionless form of the natural frequency is 
\begin{equation} \label{eq:ndnatfreq}
\omega = \sqrt{\frac{8(1+{\alpha}_2 {\Omega}^2) - 3 {\alpha_1}{\Omega}^2 (4 x_0^2 + x_1^2)}{4(2 + {\alpha}_1(2x_0^2 + x_1^2))}}.
\end{equation}
With assumptions of zero static deflection $(x_0=0)$ and small vibration amplitude $(x_1 \approx 0),$ we simplify Eq.\eqref{eq:ndnatfreq} as
\begin{equation*} \label{eq:znatfreq}
\omega = \sqrt{1 + {\alpha}_2 {\Omega}^2}.
\end{equation*}
Referring to Fig.\ref{fig:znatfreq}, the dashed curve shows variation in the natural frequency as a function of $\Omega$ mainly due to centrifugal stiffening. The primary resonance ($\Omega = \omega$) is absent for the chosen parameter values (Table \ref{table1}) in case of out-of-plane vibrations model.
\subsection{Frequency response using harmonic balance}
\label{sec:HBNC}
Method of harmonic balance is routinely applied to approximate periodic solutions of differential equations by assuming the solution to be a truncated Fourier series. It is a special case of Bubnov-Galerkin scheme of the methods of weighted residual where weighting functions are the same as shape functions.
Here, we approximate the steady-state solution to Eq.\eqref{eq:mroe} by
\begin{equation} \label{eq:fousol}
x(t) \approx A_1 \sin (\Omega t) + B_1 \cos (\Omega t) + 
A_2 \sin (2\Omega t) + B_2 \cos (2\Omega t) + 
A_3 \sin (3\Omega t) + B_3 \cos (3\Omega t) + C.
\end{equation}
Using Eq.\eqref{eq:fousol}, super-harmonic resonances up to third order can be captured. Substituting the above in Eq.\eqref{eq:mroe} and expanding, we obtain the residual in the form
\begin{align*}
\kappa_0 + \kappa_1 \sin (\Omega t) + \kappa_2 \cos (\Omega t) + 
\kappa_3 \sin (2\Omega t) + \kappa_4 \cos (2\Omega t) +  
\kappa_5 \sin (3\Omega t) + \kappa_6 \cos (3\Omega t) + \text{higher harmonics} ,
\end{align*}
where $\kappa_0$ to $\kappa_6$ are functions of $A_1$ to $A_3,$ $B_1$ to $B_3,$ and $C.$
Taking weights same as the shape functions, i.e., $\sin(\Omega t),\cos(\Omega t),\cdots, \cos(3\Omega t)$ and $\cos(0 t)=1$ and then integrating the weighted residual over one time period, i.e., $2\pi,$ we have a system of nonlinear algebraic equations that needs to be solved for different values of $\Omega.$
\begin{figure}[b!]
	\centering
	\begin{subfigure}[b]{0.49\textwidth}
		\includegraphics[width=\textwidth]{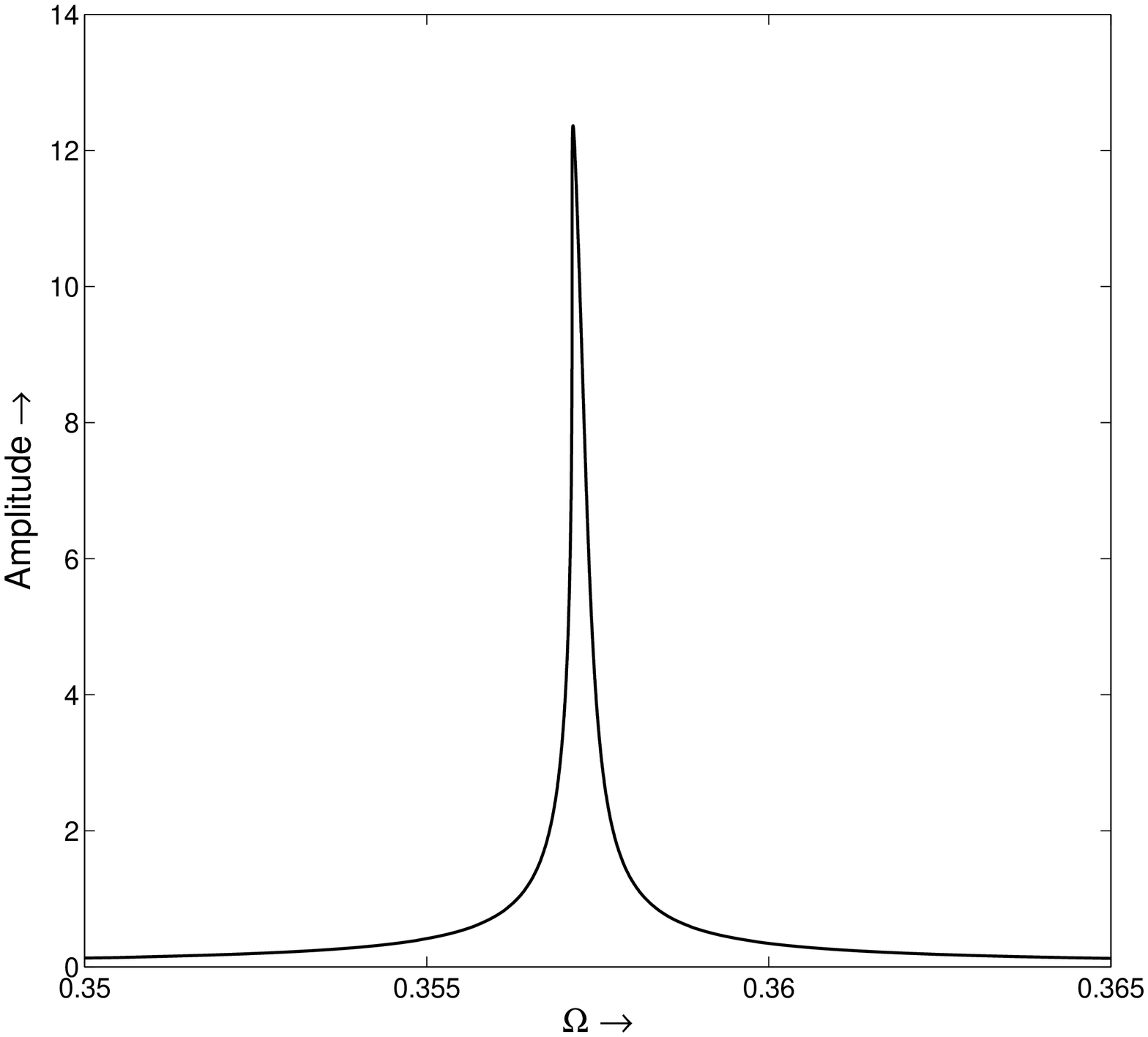}%
		\caption{}
		\label{fig:hb3o}
	\end{subfigure}
	\hfill
	\begin{subfigure}[b]{0.49\textwidth}
		\includegraphics[width=\textwidth]{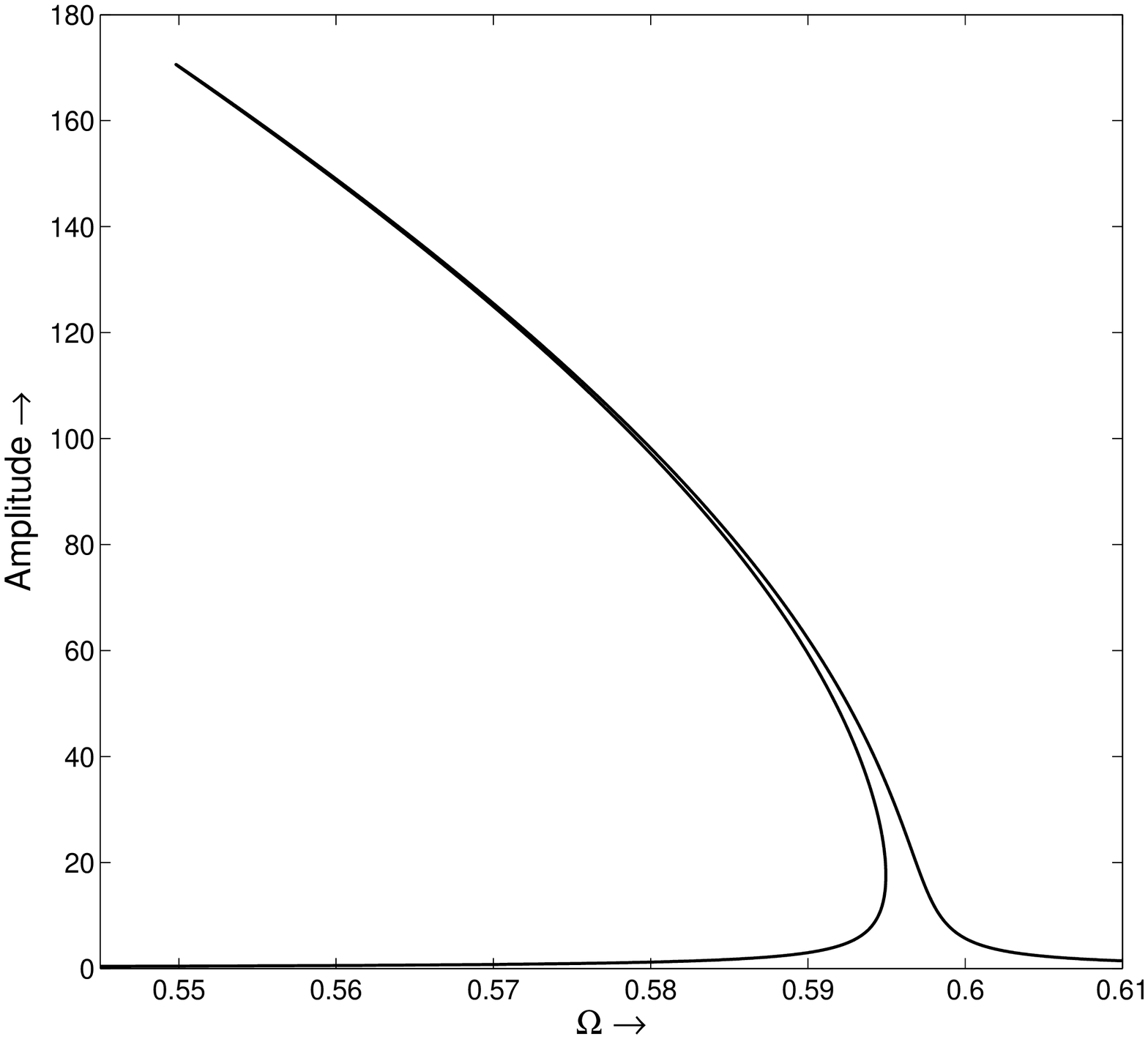}%
		\caption{}
		\label{fig:hb2o}
	\end{subfigure}
	\caption{Frequency response for Eq.\eqref{eq:mroe} using harmonic balance with parameter values from Table \ref{table:dimpar1}. (a) Tertiary resonance at $\Omega=0.3579$ (b) secondary resonance at $\Omega = 0.5969.$}
\end{figure}

Arc-length based continuation is a method to obtain approximate solutions to a system of nonlinear algebraic equations. Using this method with $\Omega$ as a continuation parameter for the above system of equations, we solve for
\begin{equation*}
\vec{V} = \begin{bmatrix}
\Omega &A_1 &B_1 &A_2 &B_2 &A_3 &B_3 &C
\end{bmatrix}^T.
\end{equation*} 
Two initial solutions $\vec{V_1}$ and $\vec{V_2}$ are found by choosing two nearby values of $\Omega$, say $\Omega_1$ and $\Omega_2$. With  $\Omega_1$ known, we solve the system of seven equations in seven unknowns using Newton-Raphson method by providing a suitable initial guess. The second solution corresponding to $\Omega_2$ is found similarly with solution corresponding to $\Omega_1$ serving as an initial guess. The nearness of the two $\Omega$ values decides the value of arc-length parameter which in turn decides how sharply the one-parameter solution curve to be obtained turns around. To find a solution $\vec{V_{j}}$ using a known solution $\vec{V_{j-1}}$, we introduce a new equation 
\begin{equation} \label{eq:arclengtheq}
||\vec{V_j} - \vec{V_{j-1}}|| = s.
\end{equation}
We now have a system of eight equations in eight unknowns. Newton-Raphson method may be used iteratively to obtain consequent solutions $\vec{V_j}$'s by providing initial guesses obtained through linear extrapolation of previous two solutions, i.e, $2\vec{V_{j-1}} - \vec{V_{j-2}}$. 

Figs.\ref{fig:hb3o} and \ref{fig:hb2o} respectively show the superharmonic resonances of the third and second order, i.e, when $3 \Omega = \omega$ and $2 \Omega = \omega$ respectively. Amplitude corresponding to the $n$th superharmonic is given by $\sqrt{A_n^2 + B_n^2}.$ 
The value of $\Omega$ at which the $n$th superharmonic resonance occurs is found by solving
\begin{equation*}
n \Omega = \omega = \sqrt{1 + {\alpha}_{2} \Omega^2} .
\end{equation*}
Resonance peak corresponding to $2 \Omega = \omega$ exhibits a soft-spring characteristic (bending toward left). 
\subsection{Frequency response using method of multiple scales}

Perturbation methods are applied to get solutions of a differential equation that contains a small parameter. To apply the method of multiple scales to Eq.\eqref{eq:mroe}, we need identify a small parameter say $\epsilon,$ with the usual assumption $0<\epsilon \ll 1$ and then accordingly order various coefficients involved (refer to Table \ref{table:dimpar1} for numerical values) in terms of the small parameter. Our choice of $\epsilon$ here is ${\alpha}_1^{1/4}=(7.1824 \times 10^{-6})^{1/4} \approx 0.0518.$ With this choice, we take ${\alpha}_2,\;{Q}_c$ and $\Delta{Q}$ to be $\mathcal{O}(1).$ Other parameters in Eq.\eqref{eq:mroe} can readily be ordered as follows
\begin{equation} \label{eq:scaling}
{\alpha}_1 \sim \mathcal{O}(\epsilon^4) ,\qquad c \sim \mathcal{O}(\epsilon^2), \qquad {g} \sim \mathcal{O}(\epsilon).
\end{equation}
Introducing $\omega$ to be natural frequency of the unperturbed oscillator, ordering Eq.\eqref{eq:scaling} into Eq.\eqref{eq:mroe} and after rearrangement of terms, we get
\begin{equation} \label{eq:eq_mu}
\ddot{x} + \left(\omega^2-\epsilon g \sin (\Omega t)\right) x
+ \epsilon^2 c\dot{x} + \epsilon^4\left( x^2\ddot{x} + \dot{x}^2 x - \frac{1}{2} \Omega^2  x^3\right) = Q_c + \Delta Q \sin (\Omega t),
\end{equation}
with 
$\;\omega^2 = 1 + \alpha_2  \Omega^2$ and $c, \;g$ of $\mathcal{O}(1).$ Nonlinearities of type cubic are relatively weakest. Also the damping term is weaker than the parametric excitation term.
Corresponding to numerical values of parameters used by Inoue et al.\cite{Inoue} in their physical experiments (Table \ref{table1}), parameters in Eq.\eqref{eq:eq_mu} take values as can be seen from Table \ref{table:dimpar2}.
\begin{table}[h]
	\centering
	\begin{tabular}{|c|c|}
		\hline
		Parameter & Value \\
		\hline
		$\epsilon$ & $0.0518$\\
		$\alpha_2$ & $1.1934$\\
		$c$ & $0.1481$		\\
		$g$ & $1.8204$		\\
		$Q_c$	& $3.73$	\\
		$\Delta Q$ & $1.24$ \\
		\hline
	\end{tabular}
	\caption{Dimensionless parameters in Eq.\eqref{eq:eq_mu}.}
	\label{table:dimpar2}
\end{table}

To study the steady-state amplitude response of the system near secondary and tertiary resonances, we introduce an $\mathcal{O}(1)$ detuning parameter $\sigma$ to express the nearness of the forcing frequency $\Omega$ to the $n$th-order superharmonic as \cite{Nayfeh}
\begin{equation} \label{eq:detuning}
\Omega = \frac{\omega + \epsilon \sigma}{n} .
\end{equation}
Since Eq.\eqref{eq:eq_mu} contains nonlinear terms up to $\mathcal{O}(\epsilon ^4)$, we wish to carry out multiple scales analysis up to $\mathcal{O}(\epsilon ^4).$ As is usual with multiple scales analysis, we assume $x(t)$ to be function of time scales $T_0, T_1, T_2, T_3$ and $T_4$ defined as		
\begin{equation} \label{eq:xT}
x(t) \equiv  X(T_0, T_1, T_2, T_3, T_4), \quad\mbox{with}\quad
\end{equation}
\begin{equation} \label{eq:Tscales}
T_0 = t, \quad T_1 = \epsilon t, \quad T_2 = \epsilon^2 t, \quad T_3 = \epsilon^3 t, \quad T_4 = \epsilon^4 t,
\end{equation}
where $T_0$ is the fast time scale and $T_1,\;T_2,\cdots$ are slow time scales.
We further expand $X(T_0, T_1, T_2, T_3, T_4) = X(T_0, \cdots ,T_4)$ as
\begin{align*} 
X(T_0, \cdots ,T_4) & = X_0(T_0, \cdots , T_4) %
+ \epsilon X_1(T_0, \cdots , T_4) + \epsilon^2 X_2(T_0, \cdots , T_4)%
\\ & + \epsilon^3 X_3(T_0, \cdots , T_4) + \epsilon^4 X_4(T_0, \cdots , T_4) + \mathcal{O}(\epsilon^5) .
\numberthis \label{eq:Xexp}
\end{align*}
First and second derivatives with respect to $t$ are respectively as follows
\begin{equation} \label{eq:dt}
\frac{d(.)}{dt} = \frac{\partial (.)}{\partial T_0} + \epsilon \frac{\partial (.)}{\partial T_1} + \epsilon^2 \frac{\partial (.)}{\partial T_2} + 
\epsilon^3 \frac{\partial (.)}{\partial T_3} + \epsilon^4 \frac{\partial (.)}{\partial T_4} + \mathcal{O}(\epsilon^5) ,
\end{equation}
\begin{align*} 
\frac{d^2 (.)}{d t^2} &= {\frac {\partial^2 (.)}{{\partial}{T_{{0}}}^{2}}} + 2 \epsilon \left( {\frac {\partial^2 (.)} {{\partial}T_{{1}} {\partial}T_{{0}}}} \right)+ {\epsilon}^{2} \left( 2 {\frac{\partial^2(.)}{{\partial}T_{{2}}{\partial}T_{{0}}}}+{\frac {\partial^2 (.)}{{\partial}{T_{{1}}}^{2}}} \right)\\
& + 2 {\epsilon}^{3} \left( {\frac {\partial^2 (.)}{{\partial}T_{{3}}{\partial}T_{{0}}}} + {\frac {\partial^2 (.)}{{\partial}T_{{2}}{\partial}T_{{1}}}} \right) + 
{\epsilon}^{4} \left( 2   {\frac{\partial^2(.)}{{\partial}T_{{4}} {\partial}T_{{0}}}} +2 {\frac {\partial^2 (.)}{{\partial}T_{{3}}{\partial}T_{{1}}}}+{\frac{\partial^2 (.)}{{\partial}{T_{{2}}}^{2}}} \right) + \mathcal{O}(\epsilon^5) .
\numberthis \label{eq:ddt}
\end{align*}
Substituting Eqs. \eqref{eq:xT} through \eqref{eq:ddt} in Eq.\eqref{eq:eq_mu}, expanding and collecting terms, we get equations at various orders of $\epsilon$ as follows
\begin{subequations}
	\begin{align*}
	\mathcal{O}(1) : \frac{\partial^2 X_0}{\partial T_0^2} + \omega^2 X_0 & =  \Delta Q \sin (\Omega T_0) + Q_c ,
	\numberthis \label{eq:O1}
	\\
	\mathcal{O}(\epsilon) : \frac{\partial^2 X_1}{\partial T_0^2} + \omega^2 X_1 & =
	g  X_{{0}} \sin( \Omega T_{{0}}) - 2 {\frac {\partial^{2} X_{{0}}}{\partial T_{{0}}\partial T_{{1}}}} ,
	\numberthis \label{eq:Omu}
	\\
	\mathcal{O}(\epsilon^2) : X_{2,00} + \omega^2 X_2 & = 
	g  X_{{1}} \sin( \Omega T_{{0}})
	- c X_{0,0}
	- X_{0,11} - 2 \big( X_{1,01} + X_{0,02} \big) ,
	\numberthis \label{eq:Omu2}
	\end{align*}
	where we introduce compact notation  $ \displaystyle X_{i,jk} = \frac{\partial^{2} X_i}{\partial T_j \partial T_k} ,$
	\begin{align*}
	\mathcal{O}(\epsilon^3) : X_{3,00} + \omega^2 X_3 = &
	g X_{{2}} \sin( \Omega T_{{0}})  
	- c \big( X_{1,0} + X_{0,1} \big)
	- X_{1,11}
	\\ & 
	- 2 \big( X_{2,01} + X_{1,02} + X_{0,03} + X_{0,12} \big) ,
	\numberthis \label{eq:Omu3}
	\\
	\mathcal{O}(\epsilon^4) : X_{4,00} + \omega^2 X_4 = &
	g  X_{{3}} \sin( \Omega T_{{0}})
	- c \big( X_{2,0} + X_{1,1} + X_{0,2} \big)
	- X_{2,11} - X_{0,22}
	\\ & 
	- 2 \big( X_{3,01} + X_{2,02} + X_{1,03} + X_{1,12} + X_{0,13} + X_{0,04} \big)
	\\ &
	- X_0 \big( X_{0,0} \big) ^ 2
	- {X_0}^2 X_{0,00}
	+ \frac{{\Omega}^{2} X_{{0}}^{3}}{2} .
	\numberthis \label{eq:Omu4}
	\end{align*}
\end{subequations}
We note that the forcing at every order except at $\mathcal{O}(1)$ is in terms of solutions to equations at all previous orders, as usually expected in a typical perturbation method.
Since oscillator at $\mathcal{O}(1)$ is a forced one, the general solution at this order consists of solution to homogeneous part and particular solution. Hence, solution to Eq.\eqref{eq:O1} is
\begin{gather*}
X_0(T_0, \cdots , T_4) = A(T_1,\cdots , T_4) \sin (\omega T_0) + B(T_1,\cdots , T_4) \cos (\omega T_0) + \frac{Q_c}{\omega^2} - \Lambda \sin(\Omega T_0),
\\
\mbox{where} \quad
\Lambda = \frac{\Delta Q}{\Omega^2 - \omega^2}.
\numberthis  \label{eq:O1sol}
\end{gather*}
Here, $A$ and $B$ vary w.r.t. slow times $T_1$ to $T_4.$
Since we seek solution $X(T_{0},\cdots,T_{4})$ to be periodic, all $X_0(T_{0},\cdots,T_{4}),\;X_1(T_{0},\cdots,T_{4}),\cdots$ are expected to be periodic. They should not contain any unboundedly growing (i.e. secular) term. To ensure this, we need to remove resonant forcing terms (forcing frequency same as natural frequency) at each order by equating their coefficients to zero. This standard process is called the removal of secular terms.
In order to obtain the slow flow (i.e. expressions for $\dot{A}$ and $\dot{B}$) using Eq.\eqref{eq:dt}, we get partial derivatives of $A$ and $B$ w.r.t. slow times by the removal of secular terms.
Substituting Eq.\eqref{eq:O1sol} and \eqref{eq:detuning} in Eq.\eqref{eq:Omu} (equation at $\mathcal{O}(\epsilon)$), expanding and combining terms involving sines and cosines of $\omega$, and equating the coefficients of $\sin (\omega T_0)$ and $\cos (\omega T_0)$ (resonant terms) to zero, we get
\begin{subequations}
	\begin{align*}
	\frac{\partial A}{\partial T_1} = \frac{g\Lambda}{4\omega}\cos(\sigma T_1)
	\numberthis \label{eq:2AT1}
	\qquad \mbox{and} \qquad
	\frac{\partial B}{\partial T_1} = \frac{g\Lambda}{4\omega}\sin(\sigma T_1).
	\end{align*}
	Solving Eq.\eqref{eq:Omu} after removing the secular terms, we obtain solution $X_1$ at $\mathcal{O}(\epsilon)$
	\footnotesize
	\begin{align*}
	X_{{1}} = {\frac {2 g}{15{\omega}^{2}}} 
	& \Bigg\{ 
	A (T_1, \cdots, T_4) \Bigg( 5  \cos \bigg( \frac{\omega T_{{0}}-\sigma T_{{1}}}{2} \bigg) + 3 \cos \bigg( \frac{3 \omega T_{{0}}+\sigma T_{{1}}}{2} \bigg) \Bigg) 
	\Bigg. 
	\\ & 
	\Bigg. 
	- B (T_1, \cdots, T_4) \Bigg(5 \sin \bigg(\frac{\omega T_{{0}}-\sigma T_{{1}}}{2} \bigg)
	+ 3 \sin \bigg( \frac{3 \omega T_{{0}} + \sigma T_{{1}}}{2} \bigg) \Bigg) 
	\Bigg\}
	-{\frac {g}{6 {\omega}^{4}}} \Bigg( 3 {\omega}^{2} \Lambda -8 {Q_c} \sin \bigg( \frac{\omega T_{{0}}+\sigma T_{{1}}}{2} \bigg)  
	\Bigg) .
	\numberthis \label{eq:2X1}
	\end{align*}
\end{subequations}
\normalsize	
Following the above mentioned procedure at higher orders, we obtain
\begin{subequations}
	\begin{align*}
	{\frac {\partial A}{\partial T_{{2}}}} &= 
	- \frac{c }{2}A \big(T_{{1}},{\cdots},T_{{4}} \big)
	+ \frac{g^2 }{15 \omega^3} B \big(T_{{1}},{\cdots},T_{{4}} \big)
	- \frac {g \big( 3 \sigma {\omega}^{3} \Lambda  + 8 g Q_{{c}} \big)  }{24 {\omega}^{5}}\cos \big( \sigma T_{{1}} \big) ,
	\\
	{\frac {\partial B}{\partial T_{{2}}}} &= 
	- \frac{g^2 }{15 \omega^3} A \big(T_{{1}},{\cdots},T_{{4}} \big)
	- \frac{c }{2}B \big(T_{{1}},{\cdots},T_{{4}} \big) 
	- \frac {g \big( 3 \sigma {\omega}^{3} \Lambda  + 8 g Q_{{c}} \big)}{24 {\omega}^{5}} \sin \big( \sigma T_{{1}} \big),
	\\
	{\frac {\partial A}{\partial T_{{3}}}} &= 
	\frac{2 g^2 \sigma }{225 \omega^4}B \big(T_{{1}},{\cdots},T_{{4}} \big)
	+\frac {g c \Lambda \big( 4 \epsilon \sigma+\omega \big) }{24 {\omega}^{3}}\sin\big( \sigma T_{{1}} \big)
	+ {\frac {{g}^{2} \Big( 23 g\omega \Lambda -100 \sigma Q_{{c}}\Big) }{450 {\omega}^{6}}} \cos \big( \sigma T_{{1}} \big),
	\\
	{\frac {\partial B}{\partial T_{{3}}}} &= 
	- \frac{2 g^2 \sigma }{225 \omega^4} A \big(T_{{1}},{\cdots},T_{{4}} \big)
	+ {\frac {{g}^{2} \Big( 23 g\omega \Lambda -100 \sigma Q_{{c}}
			\Big) }{450 {\omega}^{6}}}\sin \big( \sigma T_{{1}} \big)
	- \frac {g c \Lambda \big(4 \epsilon \sigma+\omega \big) }{24 {\omega}^{3}} \cos
	\big( \sigma T_{{1}} \big),
	\end{align*}
	After removal of secular terms, solution $X_2$ to $\mathcal{O}(\epsilon^2)$ oscillator is
	\footnotesize
	\begin{alignat*}{2}
	&\begin{aligned}
	X_{{2}} = 
	{\frac {g}{225 {\omega}^{4}}} \Bigg\{ 
	& - A \big(T_{{1}},{\cdots},T_{{4}} \big)\Bigg( 
	4 \sigma \omega \bigg[ 25 \cos\bigg( \frac{\omega T_{{0}} - \sigma T_{{1}}}{2} \bigg) + 27 \cos \bigg(\frac{3\omega T_{{0}} + \sigma T_{{1}}}{2} \bigg) \bigg]
	- 15 g \big[ 5 \sin( \sigma T_{{1}}) - \sin( 2 \omega T_{{0}}+\sigma T_{{1}}) \big]
	\Bigg) 
	\Bigg.
	\\ & \Bigg. + 
	B \big(T_{{1}},{\cdots},T_{{4}} \big) \Bigg( 
	4 \sigma \omega \bigg[ 25 \sin\bigg( \frac{\omega T_{{0}} - \sigma T_{{1}}}{2} \bigg) + 27 \sin \bigg(\frac{3\omega T_{{0}} + \sigma T_{{1}}}{2} \bigg) \bigg]
	- 15 g \big[ 5 \cos( \sigma T_{{1}}) + \cos( 2 \omega T_{{0}}+\sigma T_{{1}}) \big]
	\Bigg) 
	\Bigg\} 
	\end{aligned}
	\\ &
	\qquad \begin{aligned}
	& + \frac{2 g^2 Q_c}{3 \omega^6}
	-{\frac {4 g \Big( g\omega\Lambda - 2 \sigma Q_{{c}} \Big)  }{9{\omega}^{5}}} \sin\bigg( \frac{\omega T_{{0}} + \sigma T_{{1}}}{2} \bigg)
	+ {\frac {2  c \big( \epsilon \sigma+\omega \big) \Lambda }{3 {\omega}^{2}}} \cos \bigg( \frac{\omega T_{{0}} + \sigma T_{{1}}}{2} \bigg)
	- \frac{6 g^2 \Lambda}{25 \omega^4} \sin \bigg( \frac{3\big(\omega T_{{0}} + \sigma T_{{1}}\big)}{2} \bigg) .
	\end{aligned}
	\end{alignat*}
\end{subequations}
\normalsize
and similarly solution $X_3$ at $\mathcal{O}(\epsilon^3)$ is obtained. The procedure is again applied to Eq.\eqref{eq:Omu4} to get $\displaystyle \frac{\partial A}{\partial T_4}, \frac{\partial B}{\partial T_4}$ (\ref{appendix:equations}) and $X_4$. For brevity, expressions for $X_3$ and $X_4$ have been excluded.


Having obtained the partial derivatives of $A$ and $B$ with respect to the slow time scales, their time derivatives can now be found using Eq.\eqref{eq:dt}.
We change variables from $A(t)$ and $B(t)$ to $R(t)$ and $\beta(t)$ as
\begin{equation} \label{eq:Rbeta}
A(t) = R(t)\sin\big(\beta(t)\big) \qquad \mbox{and} \qquad B(t) = R(t)\cos\big(\beta(t)\big)
\end{equation}
and obtain equations $\dot{R}(t),\;\dot{\beta}(t)$ from $\dot{A}(t)$ and $\dot{B}(t)$ which are non-autonomous. 
Simplifying the expressions for $\dot{R}(t)$ and $\dot{\beta}(t)$ further by introducing $\phi = \beta + \sigma\epsilon t$ (as done in \cite{Nayfeh}) and using Eq.\eqref{eq:detuning}, we obtain autonomous slow flow
\begin{align}
{\frac{dR}{{d}t}} = \sum_{i=1}^{4} \epsilon^i \mathcal{R}_i(R, \phi, \Omega)
\quad \mbox{and} \quad
{\frac{d \phi}{{d}t}} = \sum_{j=0}^{4} \epsilon^j {\Phi}_j(R, \phi, \Omega) .
\numberthis \label{eq:Rphi}
\end{align}
Expressions $\mathcal{R}_1,\ldots,\mathcal{R}_4$ and ${\Phi}_0,\ldots,{\Phi}_4$ are provided in \ref{appendix:equations}. Figure \ref{fig:amp} shows the match between numerical solution to Eq.\eqref{eq:eq_mu} 
via ode45 (built-in MATLAB integrator) and the amplitude obtained by numerically integrating the slow flow, again using ode45.
We choose forcing frequency $\Omega=0.596,$ very close to resonant frequency $0.5969.$ From Fig.\ref{fig:amp}, we observe that steady-state amplitude is close to $29.79\dots .$ MMS is able to capture transients quite well and the steady-state amplitude also.

\begin{figure} [!h]
	\centering
	\includegraphics[width=0.65\linewidth]{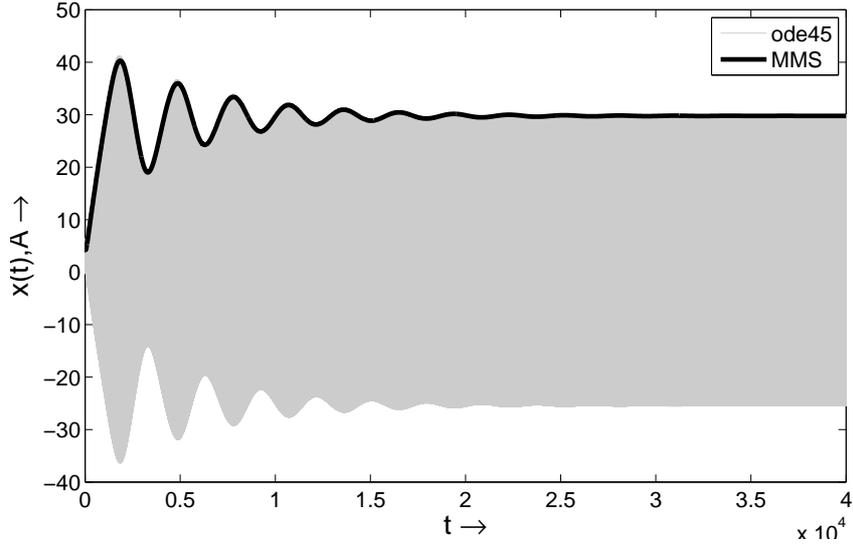}
	\caption{Amplitude via MMS and solution to Eq.\eqref{eq:eq_mu} via numerical integration for $\Omega=0.596.$ Initial conditions: $x(0)=1, \dot{x}(0)=0, R(0) = 1.8081$ and $\phi(0) = 3.5992.$}
	\label{fig:amp}
\end{figure}

\begin{figure}[h!]
	\centering
	\begin{subfigure}[b]{0.48\textwidth}
		\includegraphics[width=\textwidth]{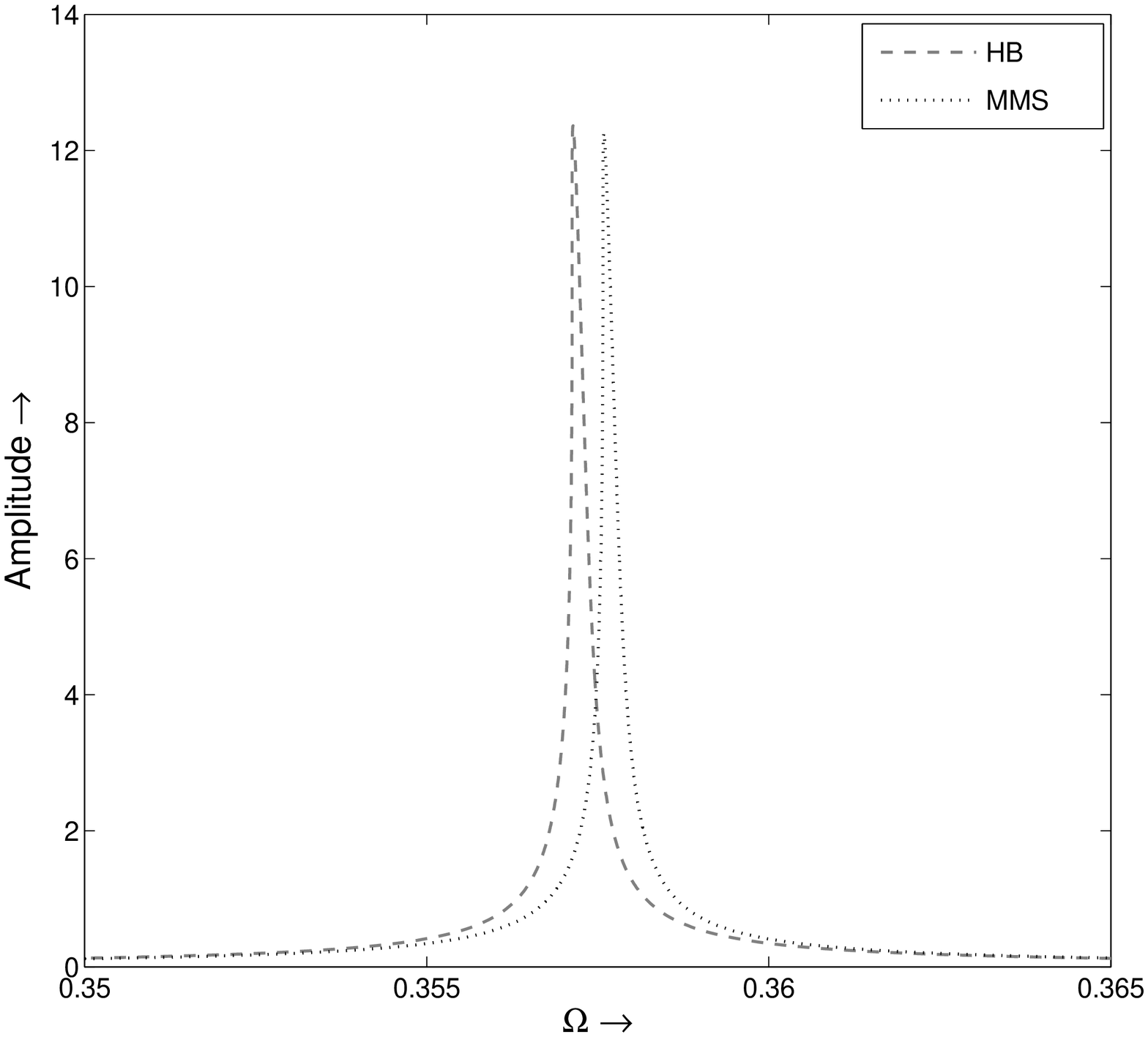}%
		\caption{}
		\label{fig:3O}
	\end{subfigure}
	\hfill
	\begin{subfigure}[b]{0.48\textwidth}
		\includegraphics[width=\textwidth]{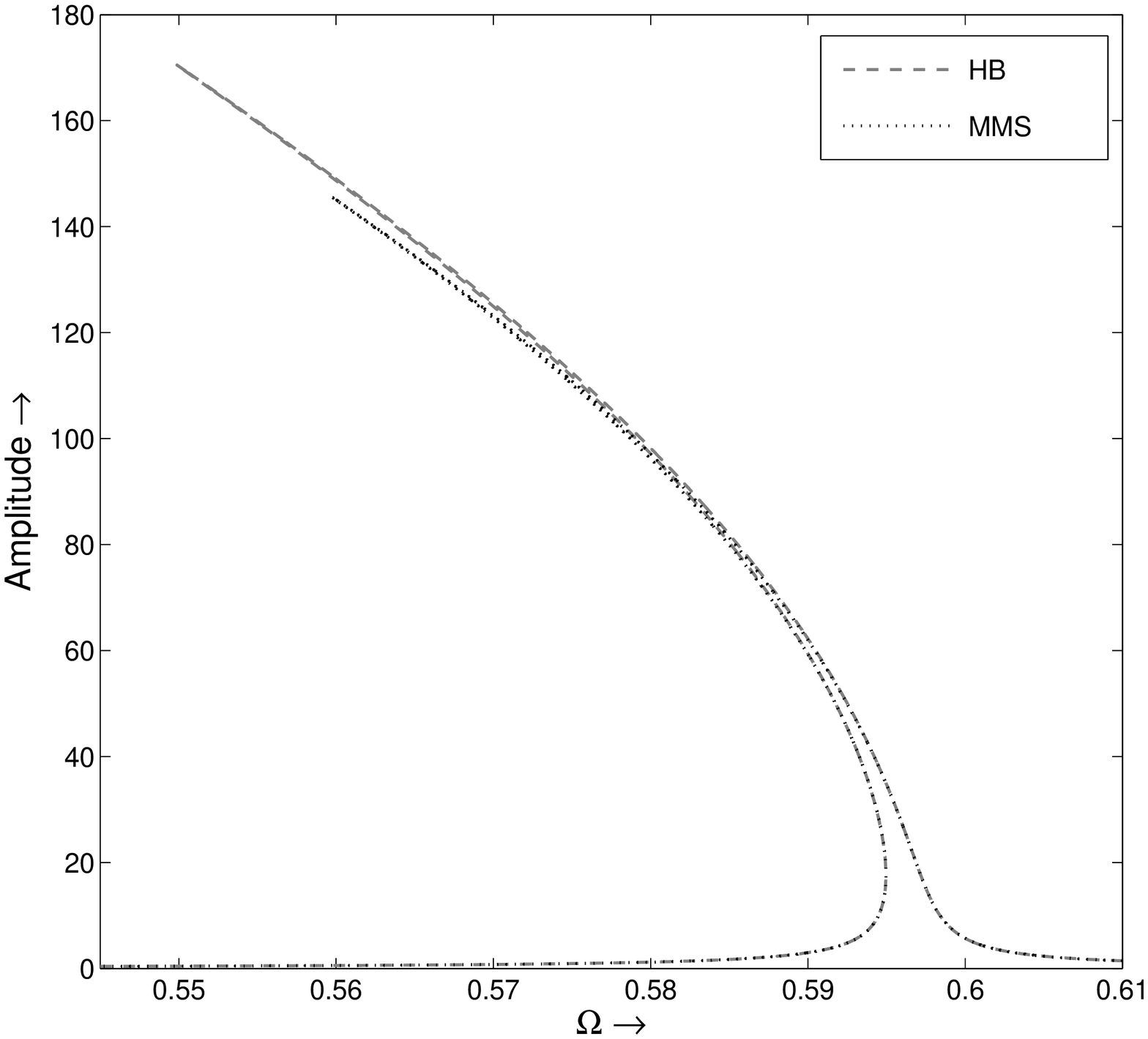}%
		\caption{}
		\label{fig:2O}
	\end{subfigure}
	\caption{Frequency responses near (a) third superharmonic and (b) second superharmonic.}
\end{figure}

In the steady-state, we have
\begin{equation*}
\dot{R} = 0 \quad \mbox{and} \quad \dot{\phi} = 0.
\end{equation*}
From Eq.\eqref{eq:Rphi}, this results in two nonlinear equations in $R$ and $\phi.$
By making $\Omega$ unknown and adding the arc-length based equation (section \ref{sec:HBNC}), we now apply continuation scheme to obtain solutions $R,$ $\phi$ and $\Omega$, and plot $R$ against $\Omega$ to get the frequency response. Figures \ref{fig:3O} and \ref{fig:2O} show the comparison between frequency response obtained via harmonic balance with continuation and the method of multiple scales near $1$:$3$ and $1$:$2$ superharmonic resonances. In case of secondary resonance (Fig.\ref{fig:2O}), match between the two frequency responses is not good for large amplitudes (of the order of 80) and the deviation increases further for larger amplitudes. However, for smaller amplitudes the match is quite good even away from the resonance. We believe that the mismatch at higher amplitudes is mainly due to imperfect ordering scheme. It is hard to come up with a better choice of numerical value of $\epsilon$ that results in a better ordering scheme. Figure \ref{fig:3O} shows frequency response via MMS that is slightly off-set with the same via harmonic balance and continuation in case of $1$:$3$ resonance $(\Omega = 0.3579).$ The off-set is of the order of $0.0005$ (against $\epsilon \approx 0.0518$).

%

\section{In-plane vibrations}
\subsection{System Description}		
We shall now derive the equation of motion for in-plane vibrations of an isolated elastic blade, proceeding with the same approach as outlined in the previous section. We assume that there are no out-of-plane and torsional motions of the blade. We here as well, model the blade as a cantilever Euler-Bernoulli beam with one end fixed to the hub. Assumption regarding the first mode-shape is the same as earlier. In case of in-plane vibrations, gravity contributes to external periodic loading in addition to the blade stiffening in the form of parametric excitation.
Aerodynamic loading  contributes to the damping and inertial forces as well as to the external periodic loading, but contributions may vary depending on the model chosen. 
\begin{figure}[h]
	\centering
	\begin{subfigure}[b]{0.48\textwidth}
		\includegraphics[width=\textwidth]{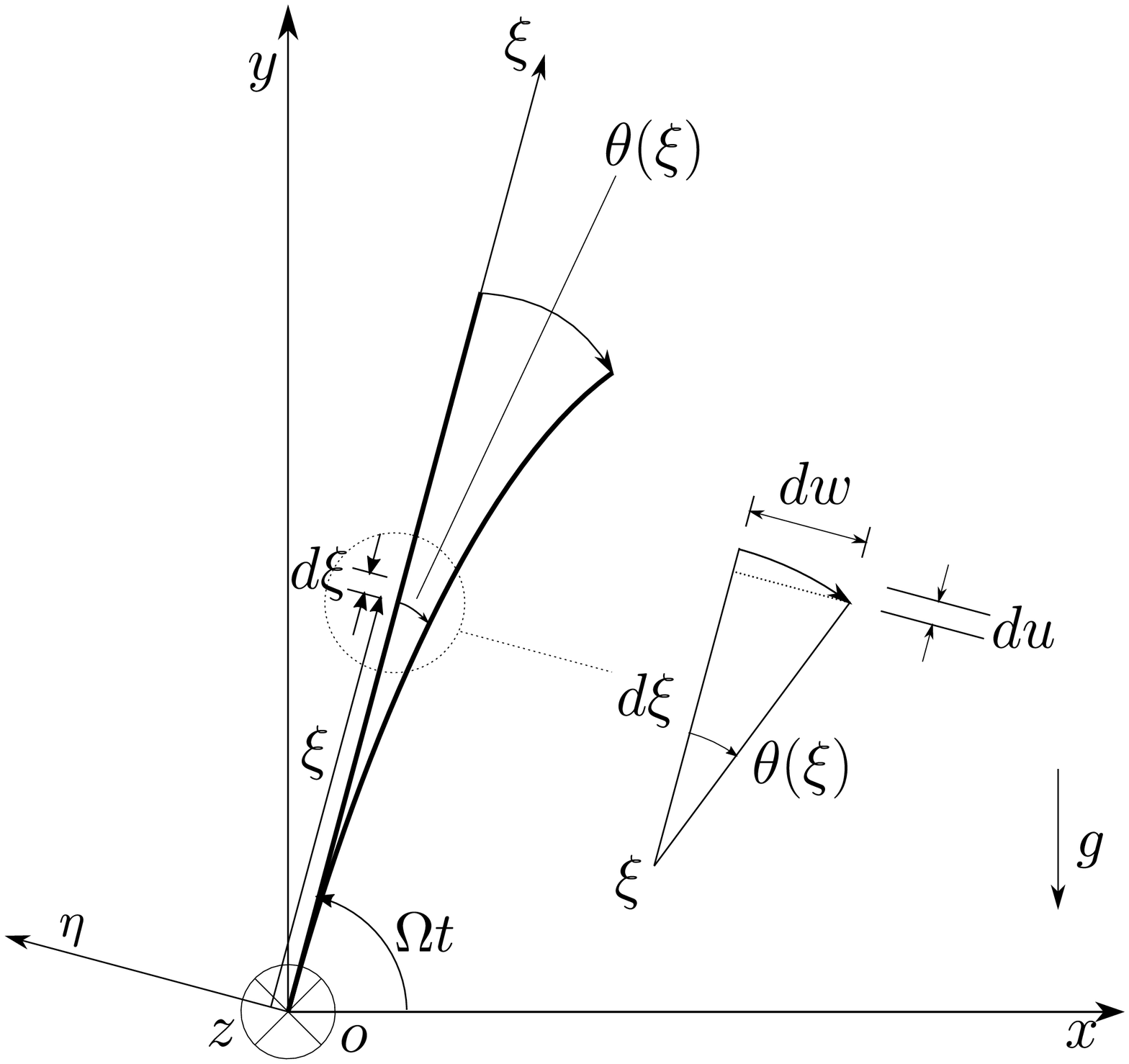}
		\caption{}
		\label{fig:in-plane}
	\end{subfigure}
	\begin{subfigure}[b]{0.4\textwidth}
		\includegraphics[width=\textwidth]{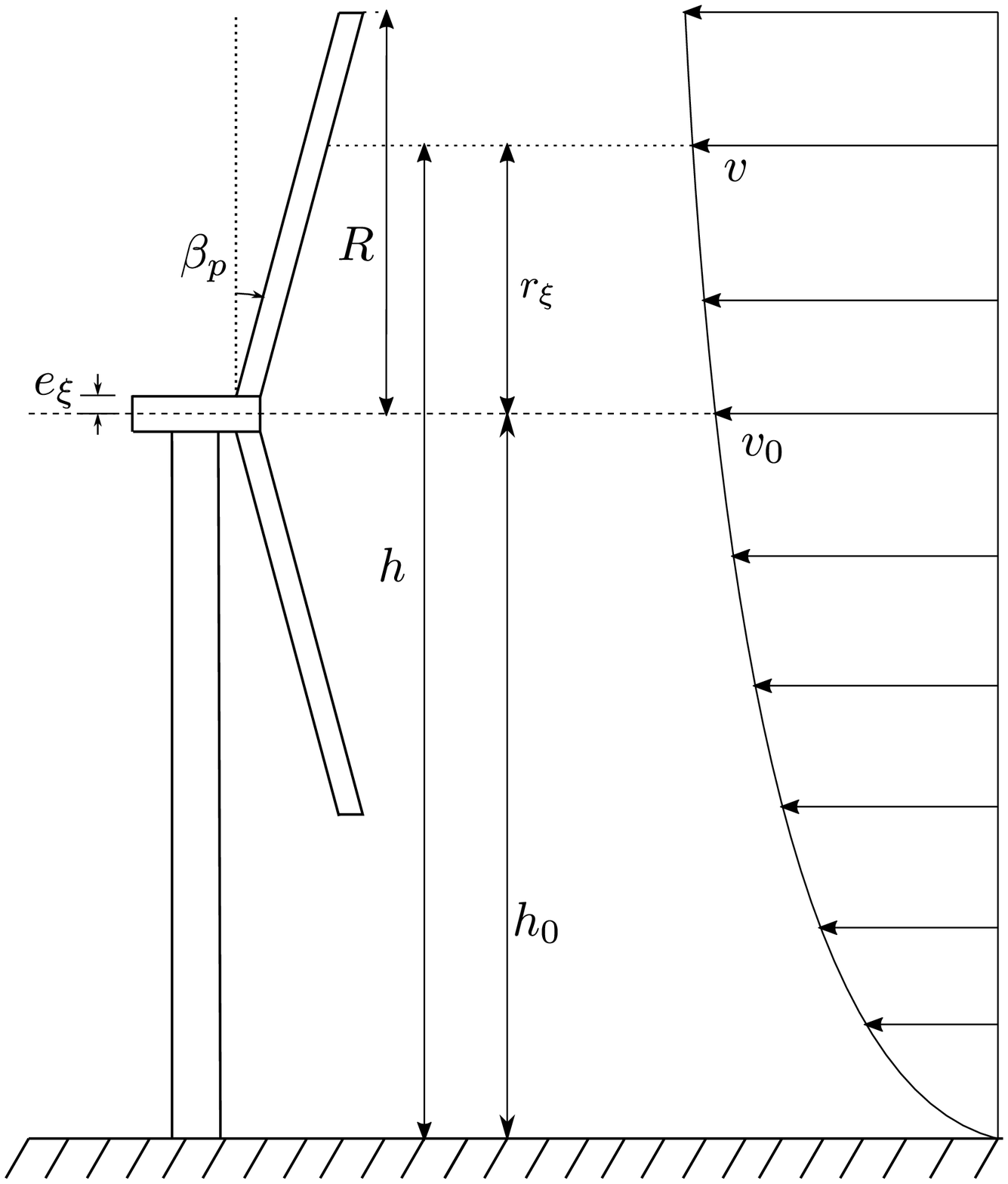}
		\caption{}
		\label{fig:wind-shear}
	\end{subfigure}
	\caption{(a) In-plane displacement of the beam (b) wind profile.}
\end{figure}
\subsubsection{Lagrangian and equation of motion}
Beam deflection with respect to the in-plane co-ordinate system $O$-$xyz$ is shown in Fig.\ref{fig:in-plane}. Let $O$-$\xi \eta z$ co-ordinate system be rigidly attached with the blade such that $\xi$-axis is aligned with $x$-axis at $t=0.$ 
As before, consider a material point on the beam at $\xi$ position. Let its co-ordinates be $(\xi, 0, 0)$ at $t=0.$ Let the same be $(x_\xi, y_\xi, z_\xi)$ at arbitrarily chosen time $t.$ Deflection of the point at time $t$ is given by
\begin{equation*}
w(\xi, t) = - \frac{\psi(\xi)}{\psi(l)} q(t) ,
\end{equation*}
where $q(t)$ is the blade-tip displacement along $\eta$-axis (Fig.\ref{fig:in-plane}) and $\psi(\xi)$ is the beam's fundamental mode-shape, given by Eq.\eqref{eq:hxi}. With the blade making an angle $\Omega t$ and deflected by an angle $\theta (\xi) = {\partial w(\xi, t)}/{\partial \xi}$, co-ordinates of the material point are given by
\begin{align} \label{eq:in-coords}
x_\xi=\big(\xi-u(\xi)\big)\cos(\Omega t) + w(\xi,t) \sin (\Omega t),
\quad
y_\xi=\big(\xi-u(\xi)\big)\sin(\Omega t) - w(\xi,t) \cos (\Omega t), 
\quad
z_\xi = 0,
\end{align}
where $u(\xi)=\alpha(\xi)q(t)^2$ with $\alpha(\xi)$ as given in Eq.\eqref{eq:uxi}.

Substituting Eq.\eqref{eq:in-coords} in Eq.\eqref{eq:TUexp}, we obtain expression for kinetic energy of the blade as
\begin{gather*}
T = \frac{\rho A}{2} 
\bigg( \frac{{\Omega}^{2}{l}^{3}}{3}
+ 2 \alpha_{{5}}\Omega \dot{q}
+ \alpha_{{3}} \dot{q}^{2}
+ \Big( 
{\Omega}^{2} (\alpha_{{3}} - 2 \alpha_{{2}})
+ 2  \alpha_{{4}}\Omega \dot{q}
+ 4 \alpha_{{1}}\dot{q}^{2} 
\Big) q^{2} 
+ \alpha_{{1}}{\Omega}^{2}q^{4}
\bigg) ,
\\
\mbox{where} \quad
\alpha_1=\int_{0}^{l}\alpha^2 d\xi,\;
\alpha_2=\int_{0}^{l}\xi \alpha d\xi,\;
\alpha_3=\frac{1}{\psi^2(l)}\int_{0}^{l}\psi^2d\xi\;,
\alpha_4=\frac{1}{\psi(l)}\int_{0}^{l}\alpha \psi d\xi\;,
\alpha_5=\frac{1}{\psi(l)}\int_{0}^{l}\xi \psi d\xi.
\numberthis \label{eq:in-Texp}
\end{gather*}
Similarly we obtain potential energy as
\begin{gather*}
U =\; \frac{1}{2}EI\beta_1 q^2 + 
g \rho A \bigg( \Big(\frac{l^2}{2}-\beta_2 q^2\Big) \sin(\Omega t)
+ \beta_3 q \cos(\Omega t) \bigg) ,
\\ \mbox{with} \quad
\begin{aligned}
\beta_1=\frac{1}{\psi(l)^2} \int_{0}^{l}\left(\frac{\partial^2{\psi}}{\partial\xi^2}\right)^2d\xi ,
\qquad
\beta_2 =\int_{0}^{l}\alpha d\xi \;
\qquad
\mbox{and}
\qquad
\beta_3 = \frac{1}{\psi(l)}\int_{0}^{l} \psi  d\xi
\end{aligned} .
\numberthis \label{eq:in-Uexp}
\end{gather*}
Rayleigh dissipation function in case of in-plane vibrations is given as
\begin{equation*}
D = \int_{0}^{l}\frac{1}{2}c_\xi(\dot{x}^2_\xi + \dot{y}^2_\xi)d\xi 
= {\frac {c_{{\xi}}}{6}} {\Big(	
	3 \big(4 \alpha_{{1}}{q}^{2}+ \alpha_{{3}} \big) {\dot{q}}^{2}
	+ 6 \big(\alpha_{{4}}{q}^{2}+\alpha_{{5}}
	\big) \Omega \dot{q}
	+3 \alpha_{{1}}{\Omega}^{2}{q}^{4}
	- 3 \big(2 \alpha_{{2}}- \alpha_{{3}} \big) {\Omega}^{2}{q}^{2}
	+{l}^{3}{\Omega}^{2}
	\Big) } ,
\end{equation*}
where $c_\xi$ is the damping coefficient per unit length. 

Aerodynamic loads depend on several factors such as local climate conditions, blade geometry, wind speed and rotational speed of the blade.
Wind shear near the earth's surface can be approximated by a power law relation as 
\begin{equation*}
\frac{v_{in}(\xi, t)}{v_0} = \Big( \frac{h}{h_0} \Big) ^p = \Big( \frac{h_0 + r_\xi \sin(\Omega t)}{h_0} \Big) ^p ,
\end{equation*}
where $v$ and $v_0$ are the velocities at heights $h$ and $h_0$ respectively, $p$ is the velocity gradient constant and $r_\xi =\xi \cos(\beta_{p})$ ($\beta_{p}$ - coning angle) is radial distance from the hub of a section at height $h$ (Fig.\ref{fig:wind-shear}).
Referring to Li et al.\cite{Li}, magnitude of distributed unsteady aerodynamic force per unit length acting on a section at $\xi$ position in $\eta$-direction (Fig.\ref{fig:in-plane}) is 
\begin{align*}
F_\eta=\frac{\rho_{{a}}ac}{2} \Bigg(& {\Omega}^{2} \bigg( {\Lambda}^{2}{R}^{2}\cos(\theta) 
- \Lambda  R \sin (\theta) \Big((1-{\beta_{{p}}}^{2})r_\xi + e_{{\xi}} \Big)
-{\frac {{r_\xi}^{2}C_{{D}}}{a}} \bigg)
\bigg.
\\& \bigg. -  \Omega \Big( \Lambda R\sin \left( \theta \right) 
+ {\frac {2 r_\xi C_{{D}}}{a
}} \bigg) {\frac {\partial w}{\partial t}}
- \frac{c \sin^2(\theta)}{2}{\frac {\partial^{2} w} {\partial {t}^{2}}}  \Bigg) ,
\numberthis \label{eq:Feta}
\end{align*}
where $\rho_a, a, c, R, e_\xi, C_D$ and $\theta$ are air density, section lift curve slope, chord length, rotor radius, hub offset, drag coefficient and blade twist respectively.
We assume that the blade has no twist $(\theta = 0)$, its root is at the center of the rotor-swept disk $(e_\xi = 0)$, and there is no coning $(\beta_p = 0, \mbox{ and therefore } r_\xi = \xi).$ We also assume chord length to be constant along the blade, same as its width $b.$ Thus, we simplify Eq.\eqref{eq:Feta} and obtain
\begin{equation}
F_\eta = \frac{\rho_{{a}} b  \Omega}{2}  \bigg( a \Omega{\Lambda}^{2}{R}^{2}
- \xi C_{{D}}(\xi \Omega + 2 {\frac {\partial  w}{\partial t}}) \bigg) .
\end{equation}
The rotor inflow ratio $\Lambda$ is defined as
\begin{gather*}
\begin{aligned}
\Lambda = \frac{v_{in}(\xi,t)}{\Omega R} 
= \Lambda_0 + \frac{r_\xi^2}{R^2}\Lambda_1 + \frac{r_\xi}{R} \Lambda_2 \sin (\Omega t) ,
\quad 	\mbox{with}
\end{aligned}
\\[0.6ex]
\begin{aligned}
\Lambda_0 = \frac{v_0}{\Omega R}, \quad
\Lambda_1 = \frac{p (p-1) R^2}{4 h_0^2}\Big(\Lambda_0 + \frac{\sigma a}{8}\Big), \quad
\Lambda_2 = -\frac{p R}{h_0}\Big(\Lambda_0 + \frac{\sigma a}{8}\Big) ,
\end{aligned}
\end{gather*}
where 
$\sigma = n_b c/(\pi R)$ is the solidity ratio for a turbine with $n_b$ number of blades.
Generalized loading due to aerodynamic forces is given by
\begin{equation}
F_a = \int_{0}^{l} F_\eta  \psi(\xi)  d\xi = 
\sum_{k=0}^{2} \Big( {Q}_k \Omega^k + {S}_k \Omega^k \sin(\Omega t) + {C}_k \Omega^k \cos(2\Omega t) \Big) 
+ {Q}_d \Omega \dot{q} = {Q}_a + {Q}_d \Omega \dot{q},
\end{equation}
where $\psi(\xi)$ is the fundamental mode-shape (Eq.\eqref{eq:hxi}). $Q_k,S_k$ and $C_k,\; k=1,2,3$ as well as $Q_d$ are constants. Here, the aerodynamic force is of harmonic excitation type. Aerodynamics also gives rise to damping term $Q_d \Omega \dot{q}.$
Defining Lagrangian in terms of kinetic and potential energies (Eqs. \eqref{eq:in-Texp} and \eqref{eq:in-Uexp}) and using Eq.\eqref{eq:Lagrange} with $Q_{ext}=F_a,$ we obtain equation of motion for in-plane vibrations as
\begin{align*} 
&\rho A\big( 4 \alpha_{{1}}{q}^{2}+\alpha_{{3}} \big)\ddot{q}
+ (c_\xi \alpha_{3} - {Q}_d \Omega) \dot{q}
- 2\rho \alpha_{{1}} A \big( {\Omega}^{2}{q}^{2} - 2 \dot{q}^{2} \big) q
+ {{4 c_{{\xi}}\alpha_{{1}}}}\dot{q}{q}^{2}
+{{c_\xi \alpha_4\Omega }}{q}^{2}	 
\\& +  \Big( EI\beta_1 + \rho A \left(  \left( 2 \alpha_{{2}}-\alpha_{{3}} \right) {\Omega}^
{2}-2 g\beta_{{2}} \sin( \Omega t) \right) \Big) q
= - g \rho \beta_{{3}} A \cos( \Omega t) 
- {{c_{{\xi}} \alpha_{{5}}\Omega }} 
+ {Q}_a .
\numberthis \label{eq:EoM-in}
\end{align*}
Numerical values of blade properties are taken from Li et al.\cite{Li} and are given in Table \ref{table:in-dimpar}.
Centrifugal stiffening is accounted for in the model. Quadratic and cubic nonlinear terms in the equation appear due to large blade deformations.

As before, we assume the solution for free vibrations of the blade with static deflection $q_0$ and free vibration amplitude $q_1$ as
\begin{equation}
\label{eq:in-freevib}
q(t) = q_0 + q_1 \sin(\omega t),
\end{equation}
where, $\omega$ is natural frequency of unforced, undamped version ($c_\xi=0, g=0, Q_d = 0$ and $Q_a = 0$) of Eq.\eqref{eq:EoM-in}. 
Substituting Eq.\eqref{eq:in-freevib} in Eq.\eqref{eq:EoM-in}, we obtain expression for natural frequency by equating the coefficient of $\sin (\omega t)$ to zero as
\begin{equation}
\label{eq:in-natfreq}
\omega = \sqrt{\frac{2 E I \beta_1 + \rho A \Omega^2 \big(4 \alpha_2 - 3 \alpha_1(4 q_0^2 + q_1^2) - 2 \alpha_3\big)}{2 \rho A \big( 2\alpha_1(2 q_0^2 + q_1^2) + \alpha_3\big)}} .
\end{equation}
Similar to the out-of-plane case, we divide Eq.\eqref{eq:EoM-in} by $b \rho \alpha_3 \omega_0^2 A,$ where $\omega_0=\sqrt{E I \beta_1/\alpha_3\rho A}$ is natural frequency in the non-rotating and small displacement condition. Introducing dimensionless versions of the variables and parameters as
\begin{gather*}
\bar{t}=\omega_0 t , \;
x ={\frac {q}{b}} \bigg(x_0 = {\frac {q_0}{b}}, \; x_1 = {\frac {q_1}{b}}\bigg), \;
\bar{\Omega}={\frac {\Omega}{\omega_{{0}}}} , \;
\bar{\beta} = \beta_2, \;
\bar{a}_{{1}}={\frac {\alpha_{{1}}{b}^{2}}{\alpha_{{3}}}}, \;
\bar{a}_{{2}}={\frac {\alpha_{{4}}b}{\alpha_{{3}}}}, \;
\bar{a}_{{3}}={\frac {\alpha_{{2}}}{\alpha_{{3}}}}, \;
\bar{a}_{{4}}={\frac {\alpha_{{5}}}{\alpha_{{3}}b}}, \;
\bar{a}_{{5}}={\frac {\beta_{{3}}}{b}}, 
\\
\bar{c}={\frac {c_{{\xi}}}{\omega_{{0}}\rho A}}, \;
\bar{g}={\frac {g}{\alpha_{{3}}{\omega_{{0}}}^{2}}}, \;
\bar{C}_{{k}}={\frac{{C}_{{k}}}{\rho \alpha_{{3}}b {\omega_{{0}}}^{2-k}A}},\;
\bar{S}_{{k}}={\frac{{S}_{{k}}}{\rho \alpha_{{3}}b {\omega_{{0}}}^{2-k}A}},\;
\bar{Q}_{{k}}={\frac{{Q}_{{k}}}{\rho \alpha_{{3}}b {\omega_{{0}}}^{2-k}A}},\;
\bar{Q}_{{d}}={\frac {{Q}_{{d}}}{\rho \alpha_{{3}} A}} ,
\mbox{ where }k=0,1,2,
\numberthis \label{eq:in-dimpar}
\end{gather*}
we get the dimensionless form of Eq.\eqref{eq:EoM-in} as
\begin{gather*}
\left( 4\bar{a}_{{1}}{x}^{2}+1 \right) \ddot{x}+4 \bar{a}_{1}x\dot{x}^{2}
+ \Big( \bar{c} \left( 4\bar{a}_{1}{x}^{2} + 1 \right) - \bar{Q}_d \bar{\Omega}\Big)\dot{x}
-2\bar{a}_1{\bar{\Omega}}^{2}{x}^{3} + \bar{a}_{{2}}\bar{c} \bar{\Omega}{x}^{2}
+ \big(  ( 2 \bar{a}_{{3}} - 1 ) {\bar{\Omega}}^{2} - 2 \bar{g} \bar{\beta}\sin( \bar{\Omega} \bar{t})  +1 \big) x
\\ = - \bar{a}_{4} \bar{c}\bar{\Omega} - \bar{a}_{{5}} {\bar{g}} \cos ( \bar{\Omega} \bar{t} ) 
+ \sum_{k=0}^{2} \Big(\bar{Q}_k \bar{\Omega}^k + \bar{S}_k \bar{\Omega}^k \sin(\bar{\Omega} \bar{t}) + \bar{C}_k \bar{\Omega}^k \cos(2\bar{\Omega} \bar{t}) \Big).
\numberthis \label{eq:ndip}
\end{gather*} 
Hereafter, we drop overbar notation representing all non-dimensional parameters. Numerical values of parameters in Eq.\eqref{eq:ndip} are given in Table \ref{table:in-ndimpar1}. Natural frequency in dimensionless form is
\begin{equation}
\omega = \sqrt{\frac{2 \big(1 + (2 a_3 - 1)\Omega^2\big) - 3 a_1 \Omega^2 (4 x_0^2 + x_1^2)}{2 \big(1 + 2 a_1 (2 x_0^2 + x_1^2) \big)}}.
\end{equation}
Assuming zero static deflection $(x_0 = 0)$ and small free vibration amplitude $(x_1 \approx 0),$ we approximate natural frequency of the system as
\begin{equation} \label{eq:in-znatfreq}
\omega = \sqrt{1 + (2a_3 - 1)\Omega^2}.
\end{equation}
\begin{figure}
	\centering
	\includegraphics[width=0.7\textwidth]{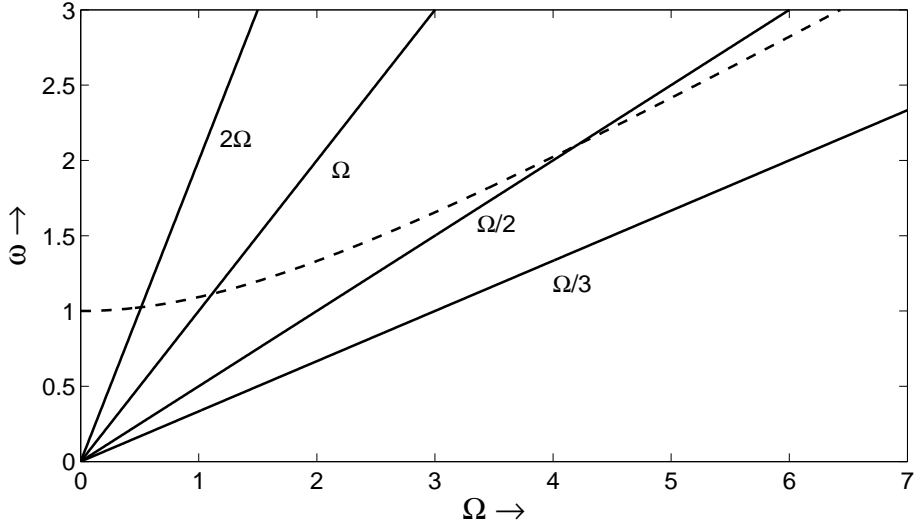}
	\caption{Natural frequency diagram}
	\label{fig:in-natfreq}
\end{figure}
Referring to Fig.\ref{fig:in-natfreq}, we see the variation of natural frequency w.r.t rotational speed $\Omega.$ As in the out-of-plane case, here also the variation is mainly due to centrifugal stiffening. Along with $1$:$1$ and superharmonic resonances, $2$:$1$ subharmonic resonance also exists due to the presence of parametric excitation and the choice of parameter values (Table \ref{table:in-dimpar}) governing the in-plane vibrations.

\subsection{Frequency response using harmonic balance}
\begin{figure}[h]
	\centering
	\begin{subfigure}[b]{0.49\textwidth}
		\includegraphics[width=\textwidth]{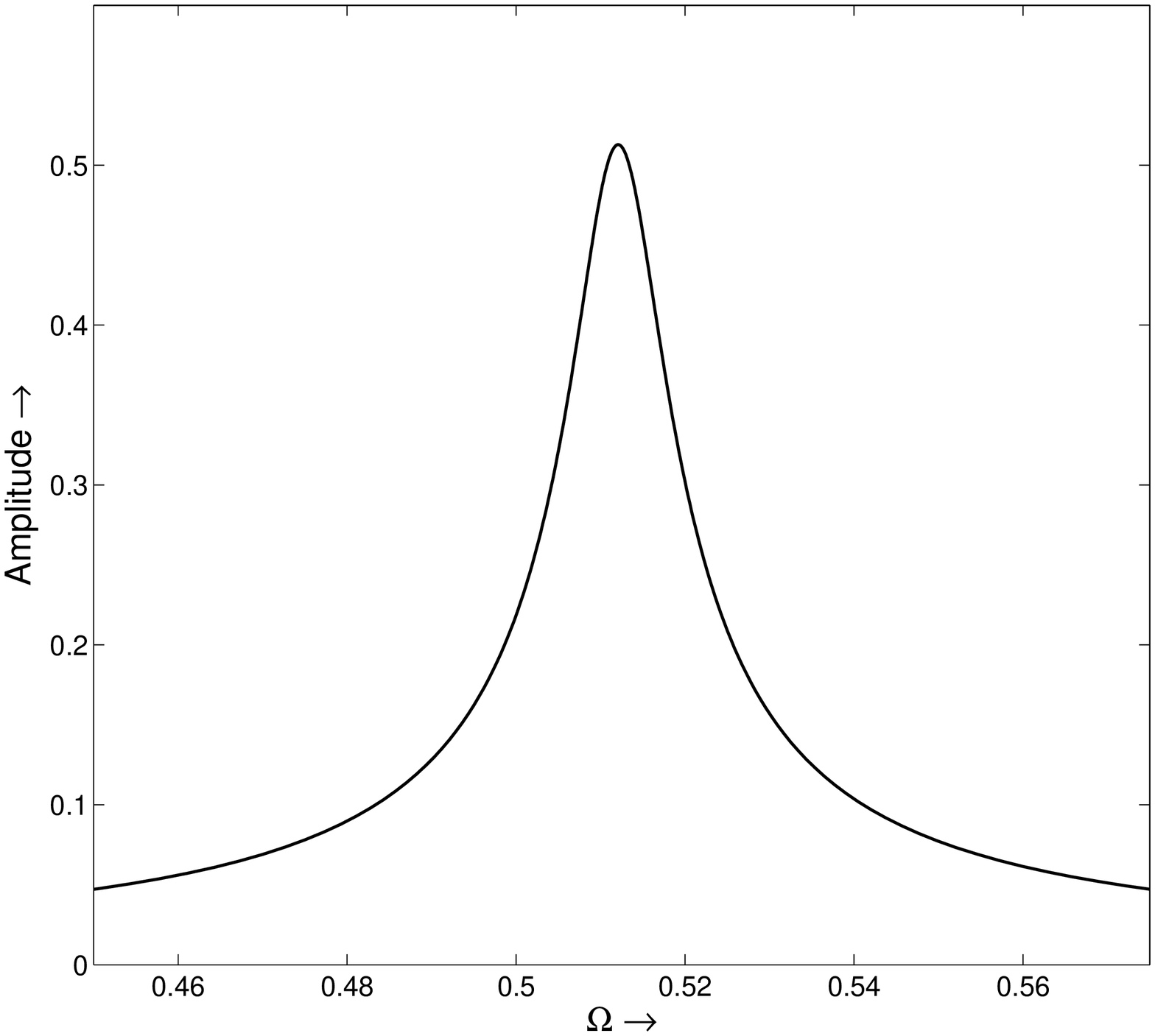}%
		\caption{}
		\label{fig:in-hb2o}
	\end{subfigure}
	\hfill
	\begin{subfigure}[b]{0.49\textwidth}
		\includegraphics[width=\textwidth]{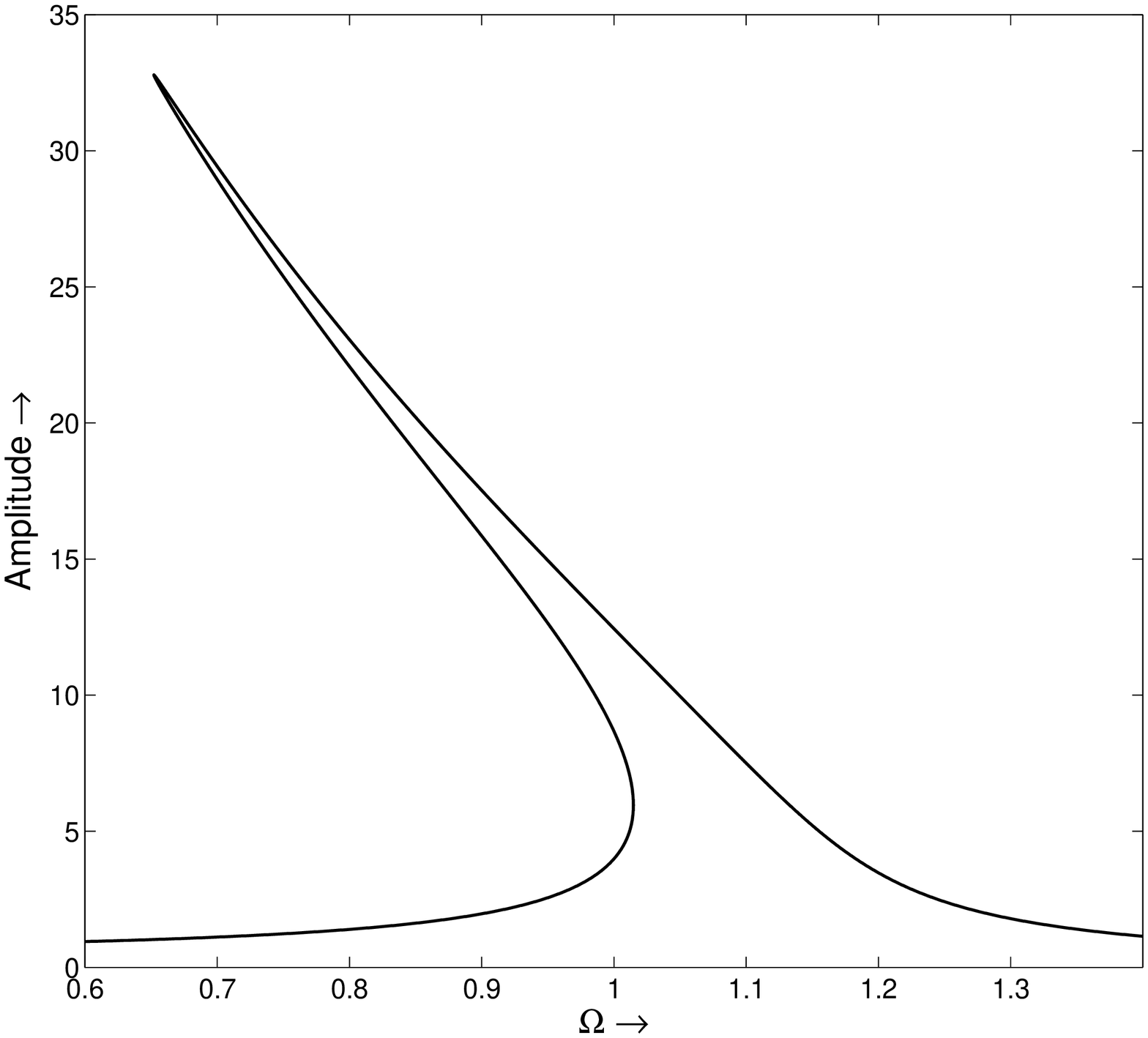}%
		\caption{}
		\label{fig:in-hb1o}
	\end{subfigure}
	\caption{Frequency response for Eq.\eqref{eq:ndip} using harmonic balance with parameter values from Table \ref{table:in-ndimpar1}. (a) Secondary resonance at $\Omega=0.5125$ (b) Primary resonance at $\Omega = 1.1134.$}
\end{figure}

As before, we obtain the frequency response to Eq.\eqref{eq:ndip} by using harmonic balance and arc-length based numerical continuation. The procedure is illustrated in section \ref{sec:HBNC}. 
Figs.\ref{fig:in-hb2o} and \ref{fig:in-hb1o} show the frequency response near secondary and primary resonances respectively. The corresponding values of $\Omega$ are obtained by solving
\begin{equation}
n \Omega = \omega = \sqrt{1 + (2 a_3 - 1)\Omega^2}. 
\end{equation} 
A soft-spring characteristic can be seen in the primary resonance due to nonlinear stiffness terms. The secondary resonance has a substantially lower peak amplitude in comparison with that of primary resonance. For the chosen numerical values of parameters, subharmonic resonance has a very low amplitude peak, which we neglect here.
\subsection{Frequency response using method of multiple scales}
\begin{figure}[h!]
	\centering
	\begin{subfigure}[b]{0.45\textwidth}
		\centering
		\includegraphics[width=\textwidth]{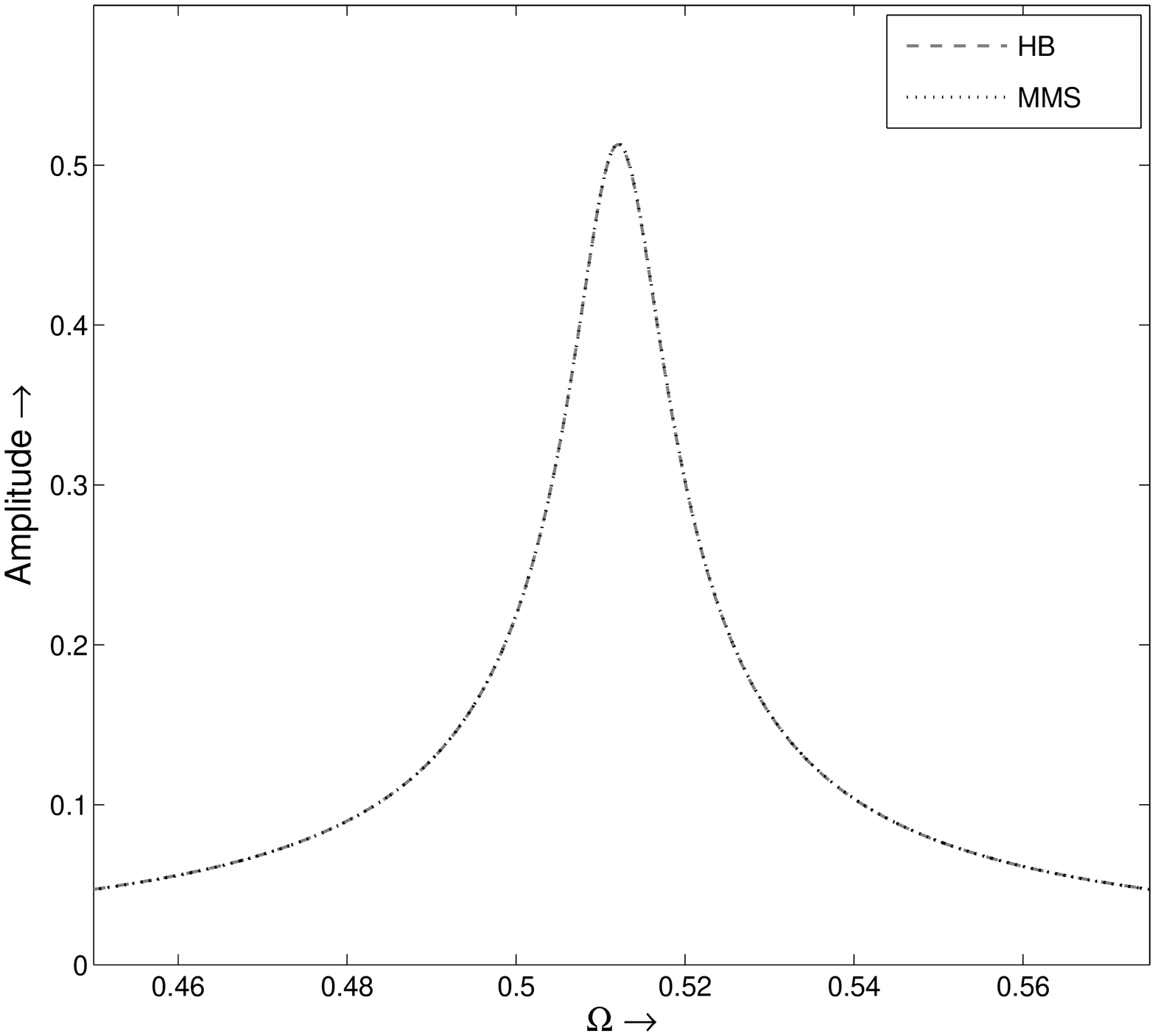}%
		\caption{}
		\label{fig:in-sec}
	\end{subfigure}
	\hfill
	\begin{subfigure}[b]{0.46\textwidth}
		\centering
		\includegraphics[width=\textwidth]{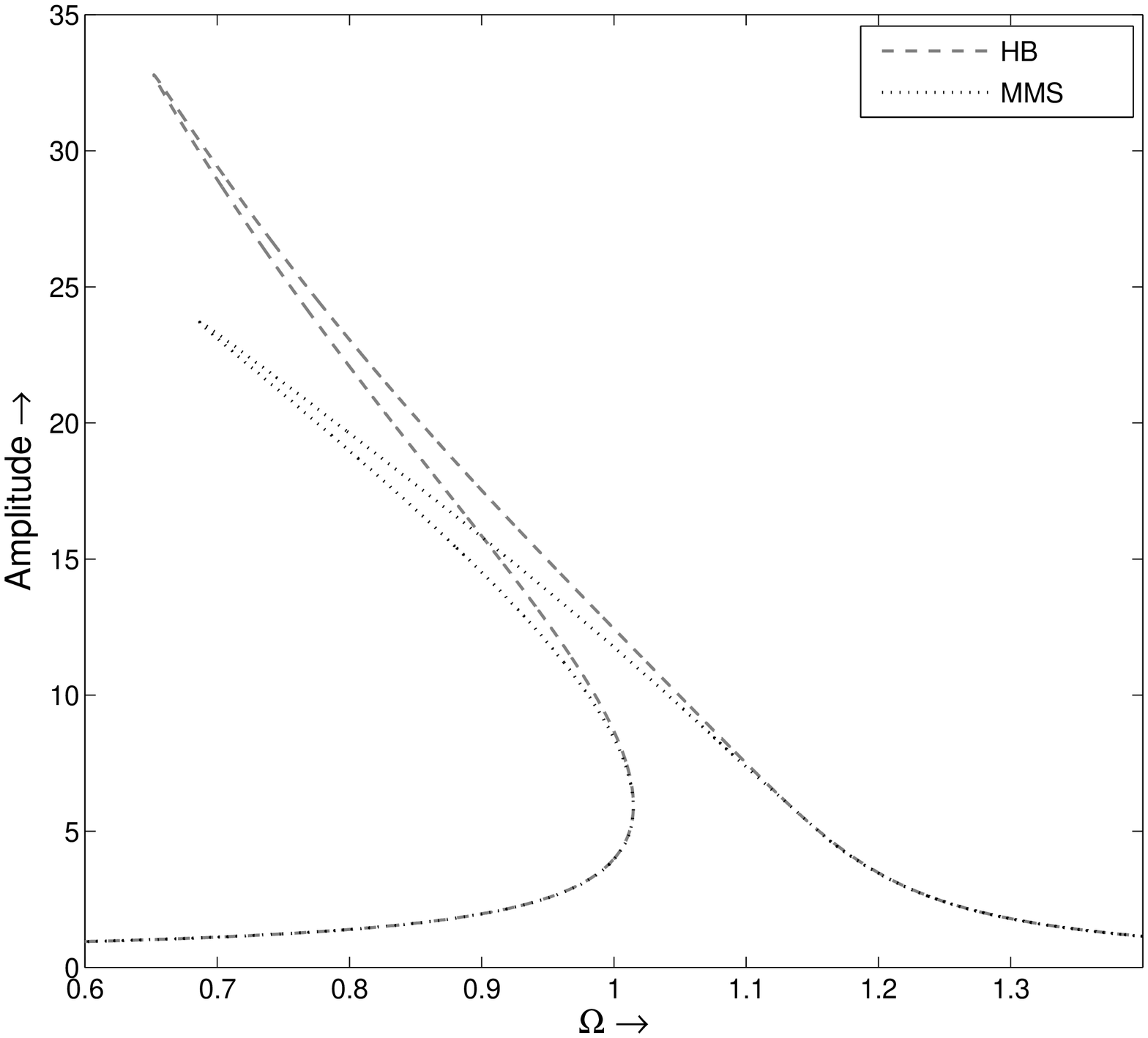}%
		\caption{}
		\label{fig:in-pri}
	\end{subfigure}
	\caption{Frequency responses near (a) second superharmonic and (b) primary resonances.}
\end{figure}

We wish to obtain frequency response in case of in-plane vibrations using the method of multiple scales for which, as mentioned previously, different coefficients involved are needed to be ordered in terms of a small parameter. According to the ordering scheme introduced in \ref{appendix:ordsch} for the choice of $\epsilon \approx 0.0215$, we rewrite Eq.\eqref{eq:ndip} governing the in-plane vibrations as 
\begin{align*}
& \ddot{x} + \omega^2 x 
+ \epsilon \big( \gamma_{1} \dot{x} + \gamma_{2}\sin(\Omega t) x
\big)
+ \epsilon^2 \big( \gamma_{3} (x^2 \ddot{x} + x \dot{x}^2) + \gamma_{4}\Omega \dot{x} + (\gamma_{5} + \gamma_{6}{\Omega}{x} ) 	\Omega{x}^{2} 
\big)
+ \epsilon^3 \gamma_{7} x^2 \dot{x} 
\\&	
= \delta_{1}\Omega + \delta_{2}\cos(\Omega t) 
+ \epsilon \big( \delta_{3} + \delta_{4}\Omega^2 + \delta_{5} \sin(\Omega t)  \big)
+ \epsilon^2 \big( \delta_{6}\Omega\sin(\Omega t) + ( \delta_{7} + \delta_{8}
\Omega) \cos \left( 2 \Omega t
\right) \big)
\\ &	\quad 
+ \epsilon^3 \Omega^2 \big(\delta_{9} \sin(\Omega t) + \delta_{10} \cos(2\Omega t) \big) ,
\numberthis \label{eq:in-meqn}
\end{align*}
where natural frequency $\omega$ of the unperturbed oscillator ($\epsilon=0$) is approximately given by Eq.\eqref{eq:in-znatfreq}.
We also introduce detuning parameter $\sigma$ to study the steady-state response near the primary and second superharmonic resonances in the same way as in Eq.\eqref{eq:detuning}.
We now carry out multiple scales analysis of Eq.\eqref{eq:in-meqn} for the primary resonance ($n = 1$) up to $\mathcal{O}(\epsilon^3).$ As done previously, we assume
\small
\begin{align}
x(t) \equiv X(T_0, \cdots ,T_3) & = X_0(T_0, \cdots , T_3) %
+ \epsilon X_1(T_0, \cdots , T_3) + \epsilon^2 X_2(T_0, \cdots , T_3)%
+ \epsilon^3 X_3(T_0, \cdots , T_3) + \mathcal{O}(\epsilon^4),
\numberthis \label{eq:in-Xexp}
\end{align}
\normalsize
with $T_i = \epsilon^i t$ for $i=0,1,2,3.$
Substituting Eq.\eqref{eq:in-Xexp} and Eqs. \eqref{eq:dt} and \eqref{eq:ddt} upto $\mathcal{O}(\epsilon^3)$ in Eq.\eqref{eq:in-meqn}, expanding and collecting terms, we obtain equations at various orders of $\epsilon$ as (using the compact notation $ \displaystyle X_{i,jk} = \frac{\partial^{2} X_i}{\partial T_j \partial T_k} $ introduced earlier)
\begin{subequations}
	\begin{align*}
	\mathcal{O}(1) : X_{0,00} + \omega^2 X_0 = &   \delta_{1} \Omega + \delta_{2} \cos (\Omega T_0) ,
	\numberthis \label{eq:in-O1}
	\\[0.5ex]
	\mathcal{O}(\epsilon) : X_{1,00} + \omega^2 X_1 = &-\Big( 2 X_{0,01} + \gamma_{1} X_{0,0} + (\gamma_{2} X_0 - \delta_{5}) \sin \big(\Omega T_0 \big)\Big) + \delta_{3} + \delta_{4} \Omega^2,
	\numberthis \label{eq:in-Oeps}
	\\[0.5ex]
	\mathcal{O}(\epsilon^2) : X_{2,00} + \omega^2 X_2 = &
	-\Big( 
	\gamma_{{3}}{X_{{0}}}^{2}X_{{0,00}} + 2(X_{{0,02}} + X_{{1,01}}) + X_{{0,11}} + \gamma_{{3}}X_{{0}}{X_{{0,0}}}^{2}
	+\gamma_{{1}}(X_{{0,1}} + X_{{1,0}}) 
	+ \gamma_{6}{\Omega}^{2}{X_{{0}}}^{3} 
	\\& \quad + \Omega(\gamma_{{4}}X_{{0,0}}+\gamma_{5}{X_{{0}}}^{2})
	+ \left( \gamma_{{2}}X_{{1}} -\delta_{{6}}\Omega \right) \sin \left( \Omega T_{{0}} \right) 
	- \left(\delta_{7} + \delta_{8} \Omega \right) \cos \left( 2 \Omega T_{{0}} \right) 
	\Big) ,
	\numberthis \label{eq:in-Oeps2}
	\\[0.5ex]
	\mathcal{O}(\epsilon^3) : X_{3,00} + \omega^2 X_3 = &
	- \Big( 
	\gamma_3 \big( ( 2 X_{{0,01}}+X_{{1,00}} ) X_{{0}}^2 
	+ 2 \big( X_{{1}}X_{0,00} + ( X_{{0,1}}+X_{{1,0}} )X_{{0,0}} \big)X_{{0}} +X_{{1}}X_{0,0}^2 \big)
	\\ & \quad + \gamma_{7} X_0^2 X_{0,0} + \gamma_{4}\Omega \big(X_{0,1} + X_{1,0} \big) 
	+ \big( X_{0,2} + X_{1,1} + X_{2,0}\big) \gamma_{1}
	\\ & \quad + 2 \big(X_{0,03}+X_{0,12}+X_{1,02}+X_{2,01} \big) +X_{{1,11}} + \big( 3 \gamma_{6} \Omega X_{{0}}+2 \gamma_{5} \big) \Omega X_0 X_1
	\\ & \quad + \big(\gamma_{2}X_2 -\delta_{9}\Omega^2 \big) \sin(\Omega T_0)
	- \delta_{10} \Omega^2 \cos(2 \Omega T_0)
	\Big),
	\numberthis \label{eq:in-Oeps3}
	\end{align*}
\end{subequations}
Solution to oscillator at $\mathcal{O}(1)$ (Eq.\eqref{eq:in-O1}) is 
\begin{gather*}
X_0 (T_0,\cdots,T_3) = A(T_1, T_2, T_3) \sin(\omega T_0) + B(T_1, T_2, T_3) \cos(\omega T_0) - \Lambda \cos(\Omega T_0) + \frac{\delta_{1}\Omega}{\omega^2}
\\
\mbox{with} \qquad \Lambda = \frac{\delta_{2}}{\Omega^2 - \omega^2} .
\numberthis \label{eq:in-O1sol}
\end{gather*}
$A$ and $B$ in Eq.\eqref{eq:in-O1sol} vary with slow time scales $T_1, T_2$ and $T_3$. As before, we wish to find partial derivatives of $A$ and $B$  by removing secular terms at various orders to obtain the slow flow ($\dot{A}$ and $\dot{B}$) using Eq.\eqref{eq:dt}. Substituting Eq.\eqref{eq:in-O1sol} in Eq.\eqref{eq:in-Oeps} and eliminating terms corresponding to resonant forcing, we obtain at $\mathcal{O}(\epsilon)$
\begin{subequations}
	\begin{align*}
	\frac{\partial A}{\partial T_1} = &
	- \frac{\gamma_{1} }{2}A(T_1,T_2,T_3)
	- \frac{{\Big( (\mu \sigma + \omega)(\gamma_{1}\omega^2\Lambda  + \gamma_{2}\delta_{1}) - \omega^2 \delta_{5}\Big)}}{2 \omega^3}\sin(\sigma T_1),
	\numberthis \label{eq:in-AT1}
	\\
	\frac{\partial B}{\partial T_1} = &
	- \frac{\gamma_{1} }{2}B(T_1,T_2,T_3)
	+ \frac{{\Big( (\mu \sigma + \omega)(\gamma_{1}\omega^2\Lambda + \gamma_{2}\delta_{1}) - \omega^2 \gamma_{6}\Big)}}{2 \omega^3}\cos(\sigma T_1)
	\numberthis \label{eq:in-BT1}
	\end{align*}
\end{subequations}
and solving thereafter, we get
\begin{align*}
X_1 = \frac{1}{6 \omega^2} \bigg( & -\gamma_{2}\Big( A ( T_{{1}},T_{{2}}, T_{{3}} )\big(\cos \left( 2 \omega T_{{0}}+\sigma T_{{1}} \right) + 3 \cos \left( \sigma T_{{1}} \right)  \big) 
-  B ( T_{{1}},T_{{2}},T_{{3}} ) \big( \sin \left( 2 \omega T_{{0}}+\sigma T_{{1}
} \right) - 3\sin \left( \sigma T_{{1}} \right)  \big) 
\\& + \Lambda \sin \left( 2 (\omega T_{{0}}
+ \sigma T_{{1}}) \right) \Big)
+ 6 \delta_{4} \left( \mu \sigma +\omega \right) ^{2}
+ 6 \delta_{{3}}
\bigg).
\end{align*}
Following the multiple scales procedure at higher orders as well, we obtain partial derivatives of $A$ and $B$ with respect to $T_2$ and $T_3$ by removing secular terms from Eqs. \eqref{eq:in-Oeps2} and \eqref{eq:in-Oeps3} respectively, and getting solutions $X_2$ and $X_3.$ As expressions for the same are too long, we do not include the same for brevity.
Changing variables from $A(t)$ and $B(t)$ to $R(t)$ and $\beta(t)$ (Eq.\eqref{eq:Rbeta}), introducing $\phi = \beta + \sigma \epsilon t$ and using Eq.\eqref{eq:detuning}, we obtain autonomous slow flow similar to the obtained in previous section
\begin{align} \label{eq:in-Rphi}
\frac{d R}{d t} = \sum_{i=1}^{3} \epsilon^i \mathcal{R}_i(R,\phi, \Omega)
\quad \mbox{and} \quad
\frac{d \phi}{d t} = \sum_{j=0}^{3} \epsilon^j {\Phi}_j(R,\phi, \Omega) .
\end{align}
Expressions $\mathcal{R}_1, \mathcal{R}_2, \mathcal{R}_3$ and ${\Phi}_0$ to ${\Phi}_3$ are provided in \ref{appendix:equations}.
Figure \ref{fig:in-amp} shows the match between solution to Eq.\eqref{eq:in-meqn} 
via ode45 and the amplitude obtained by numerically integrating the above slow flow, again using ode45.
Amplitude profile obtained by integrating the slow flow captures transient behaviour of the solution to Eq.\eqref{eq:in-meqn} quite well. The steady-state amplitude match is excellent. 
In a similar manner as before, equating the slow flow (Eqs. \eqref{eq:in-Rphi}) to zero gives two nonlinear algebraic equations in unknowns $R$ and $\phi.$ Again, taking $\Omega$ as unknown and adding arc-length based equation, we apply numerical continuation scheme (section \ref{sec:HBNC}) to obtain $R, \phi$ and $\Omega$. Frequency response is then obtained by plotting the steady-state amplitude $\big(\sqrt{R^2 \sin^2(\phi) + (R\cos(\phi)-\Lambda)^2}\big)$ against $\Omega$ for the primary resonance. For secondary resonance we plot $R$ against $\Omega.$ Figures \ref{fig:in-sec} and \ref{fig:in-pri} respectively show frequency response near secondary and primary resonances obtained by the method of multiple scales compared with that obtained using harmonic balance and continuation.	
In case of primary resonance (Fig.\ref{fig:in-pri}), the two response curves match well at lower amplitudes, even away from the resonance, but deviate for larger amplitudes. As stated earlier, a better ordering scheme may improve the match. For secondary resonance (Fig.\ref{fig:in-sec}), amplitude predicted by MMS matches exactly with that obtained using harmonic balance. 
\begin{figure} [h!]
	\centering
	\includegraphics[width=0.65\linewidth]{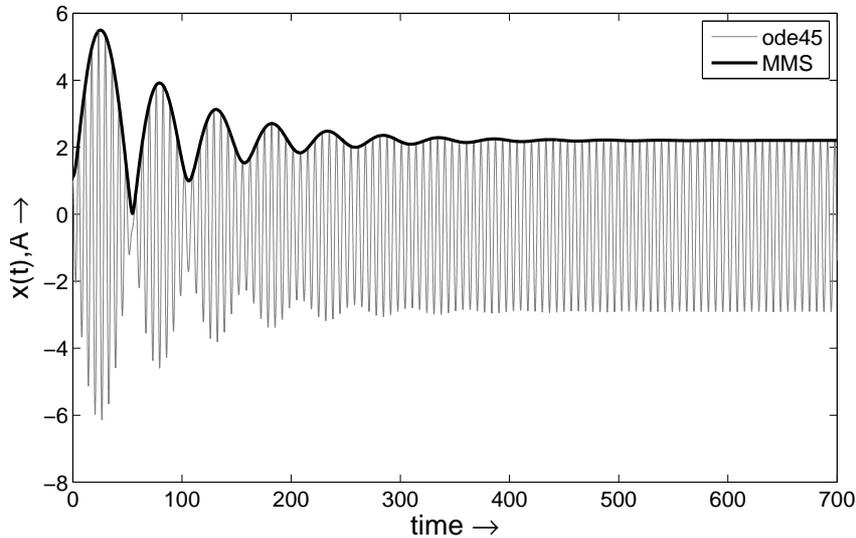}
	\caption{Amplitude via MMS and solution to Eq.\eqref{eq:ndip} via numerical integration with parameter values from Table \ref{table:in-ndimpar1} for $\Omega=0.95.$ Initial conditions: $x(0)=1, \dot{x}(0)=0, R(0)=-1.9269$ and $\phi(0)=0.$}
	\label{fig:in-amp}
\end{figure}
\section{Conclusion and future work}
We obtain frequency response curves using harmonic balance and a numerical continuation scheme in case of secondary and tertiary resonances of the out-of-plane vibrations of an isolated wind turbine blade. Ordering the governing equation in terms of a suitably chosen small parameter, we apply singular perturbation method of multiple scales. We compared frequency response curves obtained via both the above mentioned methods and find the match reasonably satisfactory. We derive the governing equation in case of in-plane vibrations of the isolated wind turbine blade following the same procedure and get the frequency response curves for various resonances.

We may further improve mathematical models for both in-plane and out-of-plane vibrations by better approximating the physical reality, for example, geometric nonlinearities may be captured more accurately by considering nonlinear curvature of the beam and by accounting for nonlinear strain. We may also assume the blade as a Timoshenko beam instead of an Euler-Bernoulli beam. Modeling of aerodynamic forces may be improved by taking into account the effects of rotor coning, blade twist, hub offset, tower shadow, tip losses, wake correction etc. Method of multiple scales may also be applied to a $2$-DOF model governing coupled out-of-plane and in-plane motions.


\appendix
\section{Ordering scheme for equation governing in-plane vibrations}
\label{appendix:ordsch}
We rewrite Eq.\eqref{eq:ndip} here for convenience
\begin{gather*}
\left( 4a_{{1}}{x}^{2}+1 \right) \ddot{x}+4 a_{1}x\dot{x}^{2}
+ \Big( c \left( 4a_{1}{x}^{2} + 1 \right) - Q_d \Omega\Big)\dot{x}
-2a_1{\Omega}^{2}{x}^{3} + a_{{2}}c \Omega{x}^{2}
+ \big(  ( 2 a_{{3}} - 1 ) {\Omega}^{2} - 2 \beta {g}\sin( \Omega t)  +1 \big) x
\\ = - a_{4} c\Omega - a_{{5}}{g} \cos ( \Omega t ) 
+ \sum_{k=0}^{2} \Big(Q_k \Omega^k + S_k \Omega^k \sin(\Omega t) + C_k \Omega^k \cos(2\Omega t) \Big).
\numberthis \label{eq:appndx-ndip}
\end{gather*} 
\begin{table}[h]
	\parbox[b]{.45\linewidth}{
		\centering
		\begin{tabular}[b]{|l|r|}
			\hline \textbf{Parameter} & \textbf{Value} \\ 
			\hline Blade length, $l$ & $48  m$ \\ 
			\hline Blade thickness, $d=0.0025l$ & $ 0.12   m$ \\ 
			\hline Blade width, $b=0.04l$ & $ 1.92   m$ \\ 
			\hline Young's modulus, $E$ & $30   GPa$ \\ 
			\hline Mass Density, $\rho$ & $1800   kg/m^3$ \\ 
			\hline Damping coefficient, $\zeta$ & $0.01 $ \\
			\hline Air density, $\rho_a$ & $1.25   kg/m^3$ \\ 
			\hline Drag coefficient, $C_D$ & $0.016$ \\ 
			\hline Hub height, $h_0$ & $ 96 m$ \\ 
			\hline Wind velocity at hub, $v_0$ & $ 12   m/s$ \\ 
			\hline Velocity gradient, $p$ & $0.167$ \\
			\hline Number of blades, $n_b$ & $3$ \\
			\hline 
		\end{tabular} 
		\caption{Parameter values of the elastic blade.}
		\label{table:in-dimpar}
	}
	\hfill
	\parbox[b]{.46\linewidth}{
		\centering
		\begin{tabular}[b]{|c|c|}
			\hline
			\textbf{Parameter} & \textbf{Value} \\
			\hline $a_1$ & $4.5967 \times 10^{-4}$\\
			\hline $a_2$ & $0.0213$\\
			\hline $a_3$ & $0.5967$\\
			\hline $a_4$ & $28.4415$\\
			\hline $a_5$ & $9.7873$\\
			\hline $\beta$ & $0.1964$\\
			\hline $c$ & $0.02$\\
			\hline ${g}$ & $0.0686$		\\
			\hline $Q_d$ & $-3.5846 \times 10^{-3}$\\
			\hline $Q_0$ & $0.1777$\\
			\hline $Q_1$ & $-4.3525 \times 10^{-4}$\\
			\hline $Q_2$ & $-0.049426$\\
			\hline $S_0$ & $-0.0216$\\
			\hline $S_1$ & $-0.0089$\\
			\hline $S_2$ & $2.0839 \times 10^{-4}$\\
			\hline $C_0$ & $-3.5512 \times 10^{-4}$\\
			\hline $C_1$ & $-2.9430 \times 10^{-4}$\\
			\hline $C_2$ & $-0.6097 \times 10^{-4}$\\
			\hline
		\end{tabular}
		\caption{Dimensionless parameters in Eq.\eqref{eq:ndip}.}
		\label{table:in-ndimpar1}	
	}
\end{table}

We now choose value of small parameter $\epsilon = 10^{-5/3} \approx 0.0215$ and order various terms appearing in Eq.\eqref{eq:appndx-ndip} in the following manner:
\begin{gather*}
4 a_{{1}} = {\epsilon}^{2}\gamma_{{3}}, \;
4 a_{{1}}c = {\epsilon}^{3}\gamma_{{7}}, \;
- Q_d = {\epsilon}^{2}\gamma_{{4}}, \;
- 2 a_{{1}} = {\epsilon}^{2}\gamma_{6}, \;
a_{{2}} c = {\epsilon}^{2}\gamma_{5}, \;
-2 {g} \beta = \epsilon \gamma_{{2}}, \;
c = \epsilon \gamma_{{1}}, \;
- a_{4} c + Q_{1} = \delta_{{1}}, \;
\\
- a_{5} {g} = \delta_{{2}}, \;
Q_{0} = \epsilon \delta_{{3}}, \;
Q_{2} = \epsilon \delta_{{4}}, \;
S_{0} = \epsilon \delta_{{5}}, \;
S_{1} = {\epsilon}^{2}\delta_{{6}}, \;
S_{2} = {\epsilon}^{3}\delta_{{9}}, \;
C_{0} = {\epsilon}^{2}\delta_{7}, \;
C_{1} = {\epsilon}^{2}\delta_{8}, \;
C_{2} = {\epsilon}^{3}\delta_{{10}} \;
\end{gather*}
with $\gamma_1$ to $\gamma_7$ and $\delta_1$ to $\delta_{10}$ are $\mathcal{O}(1).$ Introducing the above ordering scheme in Eq.\eqref{eq:appndx-ndip}, we get 
\begin{align*}
& \ddot{x} + \omega^2 x 
+ \epsilon \big( \gamma_{1} \dot{x} + \gamma_{2}\sin(\Omega t) x
\big)
+ \epsilon^2 \big( \gamma_{3} (x^2 \ddot{x} + x \dot{x}^2) + \gamma_{4}\Omega \dot{x} + (\gamma_{5} + \gamma_{6}{\Omega}{x} ) 	\Omega{x}^{2} 
\big)
+ \epsilon^3 \gamma_{7} x^2 \dot{x} 
\\&	
= \delta_{1}\Omega + \delta_{2}\cos(\Omega t) 
+ \epsilon \big( \delta_{3} + \delta_{4}\Omega^2 + \delta_{5} \sin(\Omega t)  \big)
+ \epsilon^2 \big( \delta_{6}\Omega\sin(\Omega t) + (\delta_{7} + \delta_{8}
\Omega) \cos \left( 2 \Omega t
\right) \big)
\\ &	\quad 
+ \epsilon^3 \Omega^2 \big(\delta_{9} \sin(\Omega t) + \delta_{10} \cos(2\Omega t) \big) ,
\end{align*}
where $\omega^2 = 1 + (2a_{3} - 1)\Omega^2.$

\section{Expressions}
\subsection{Out-of-plane vibrations - second superharmonic}
\label{appendix:equations}

\begin{align*}
{\frac {\partial A}{\partial T_{{4}}}} =& 
- {\frac {{g}^{4} \sin \big( 2 \sigma T_{{1}} \big)  }{{18\omega}^{7}} A \big(T_{{1}},{\cdots},T_{{4}} \big)} 
+\bigg( {\frac {19 {g}^{2}{\sigma}^{2}}{3375 {\omega}^{5}}}
+ {\frac { \big( 5 {\omega}^{4}{\Lambda}^{2} + 6 {Q_{{c}}}^{2}
		\big)\big(\epsilon\sigma + 2 \omega \big) \epsilon\sigma}{32{\omega}^{5}}}
+ {\frac {13 {\Lambda}^{2}{\omega}^{4}+4 {c}^{2}{\omega}^{2}+22
		{Q_{{c}}}^{2}}{32 {\omega}^{3}}} 
\bigg. 
\\& \bigg. + {\frac {{g}^{4} \left( 375 \cos \left( 2 \sigma T_{{1}} \right) + 137 \right) }{6750 {\omega}^{7}}} 
+ {\frac { \big( 3 {\epsilon}^{2}{\sigma}^{2} + 6 \epsilon \sigma \omega + 19 {\omega}^{2} \big)}{64 \omega} \Big({A\big(T_{{1}},{\cdots},T_{{4}} \big)}^{2} + {B\big(T_{{1}},{\cdots},T_{{4}} \big)}^{2} \Big) }
\bigg) B\big(T_{{1}},{\cdots},T_{{4}} \big)
\\& 
+ {\frac {cg \Big( 32 \epsilon {\sigma}^{2}{\omega}^{2} \Lambda +41 \sigma{\omega}^{3} \Lambda - 40 gQ_{{c}} \Big) \sin \big( \sigma T_1 \big) }{288 {\omega}^{6}}}
\\& 
+ \Big( 
{\frac {{g}^{2} \left( 22287 g\sigma{\omega}^{3} \Lambda -14000 {\sigma}^{2}{\omega}^{2}Q_{{c}}-4280 {g}^{2}Q_{{c}} \right) }{54000 {\omega}^{9}}}
-{\frac {9 {\Lambda}^{2}Q_{{c}} \left( {\epsilon}^{2}{\sigma}^{2} + 2 \epsilon \sigma \omega + {\omega}^{2} \right) }{32 {\omega}^{3}}}
\Big) \cos \big( \sigma T_1 \big)
\\
\\
{\frac {\partial B}{\partial T_{{4}}}} =& 
{\frac {{g}^{4} \sin \big( 2 \sigma T_{{1}} \big)  }{{18\omega}^{7}} B \big(T_{{1}},{\cdots},T_{{4}} \big)}
- \bigg( {\frac {19 {g}^{2}{\sigma}^{2}}{3375 {\omega}^{5}}}
+ {\frac { \big( 5 {\omega}^{4}{\Lambda}^{2} + 6 {Q_{{c}}}^{2}
		\big)\big(\epsilon\sigma + 2 \omega \big) \epsilon\sigma}{32{\omega}^{5}}}
+ {\frac {13 {\Lambda}^{2}{\omega}^{4}+4 {c}^{2}{\omega}^{2}+22
		{Q_{{c}}}^{2}}{32 {\omega}^{3}}} 
\bigg. 
\\& \bigg. + {\frac {{g}^{4} \left( - 375 \cos \left( 2 \sigma T_{{1}} \right) + 137 \right) }{6750 {\omega}^{7}}} 
+ {\frac { \big( 3 {\epsilon}^{2}{\sigma}^{2} + 6 \epsilon \sigma \omega + 19 {\omega}^{2} \big)}{64 \omega} \Big({A\big(T_{{1}},{\cdots},T_{{4}} \big)}^{2} + {B\big(T_{{1}},{\cdots},T_{{4}} \big)}^{2} \Big) }
\bigg) A\big(T_{{1}},{\cdots},T_{{4}} \big)
\\&
+ \Big( {\frac {{g}^{2} \left( 22287 g\sigma{\omega}^{3} \Lambda -14000 {\sigma}^{2}{\omega}^{2}Q_{{c}}-4280 {g}^{2}Q_{{c}} \right) }{54000 {\omega}^{9}}}
-{\frac {9 {\Lambda}^{2}Q_{{c}} \left( {\epsilon}^{2}{\sigma}^{2} + 2 \epsilon \sigma \omega + {\omega}^{2} \right) }{32 {\omega}^{3}}}
\Big) \sin \big( \sigma T_1 \big)
\\&
- {\frac {cg \Big( 32 \epsilon {\sigma}^{2}{\omega}^{2} \Lambda +41 \sigma{\omega}^{3} \Lambda - 40 gQ_{{c}} \Big) \cos \big( \sigma T_1 \big) }{288 {\omega}^{6}}}
\end{align*}

Coefficients in equation for $\dot{R}$ in
Eq.\eqref{eq:Rphi}:
\begin{subequations}
	\begin{align*}
	&\begin{aligned}
	\mathcal{R}_1 = - \frac {g \Lambda \big(2\Omega -3\omega\big)}{8\omega^2}\sin ( \phi )
	\end{aligned},
	\\[0.6ex]
	&\begin{aligned}
	\mathcal{R}_2 = -{\frac { 2g^2 Q_c \big(14\Omega^2 - 8\omega \Omega + 5 \omega^2 \big) } {27{\omega}^{7}}}\sin ( \phi)
	- \frac{c R}{2}
	\end{aligned},
	\\[0.6ex]
	&\begin{aligned}
	\mathcal{R}_3 = {\frac {23{g}^{3}\Lambda \big( 646 \Omega - 283 \omega \big)}{18000 {\omega}^{6}}} \sin ( \phi)
	-\frac {c g \Lambda \big(128\Omega^2 + 50\omega\Omega -45 \omega^2\big)}{288{\omega}^{4}}\cos( \phi )
	\end{aligned},
	\\[0.6ex]
	&\begin{aligned}
	\mathcal{R}_4 = & -{\frac {\big(428 g^4 + 6075 \omega^6 \Omega^2 \Lambda^2 \big)  Q_c}{5400 {\omega}^{9}}}
	\sin ( \phi)
	+{\frac {5 cg^2 Q_c }{36 {\omega}^{6}}} \cos( \phi )
	+ {\frac { {g}^{4} R }{18 {\omega}^{7}}}\sin \big( 2\phi \big) 			
	\end{aligned},
	\end{align*}
\end{subequations}

Coefficients in equation for $\dot{\phi}$ in Eq.\eqref{eq:Rphi}:
\begin{subequations}
	\begin{align*}
	& \bar{\Phi}_0 = 2 \Omega - \omega
	\\
	&\begin{aligned}
	\bar{\Phi}_1 = - \frac {g \Lambda \big(2\Omega -3\omega\big)}{8\omega^2 R}\cos ( \phi )
	\end{aligned},
	\\[0.6ex]
	&\begin{aligned}
	\bar{\Phi}_2 = -{\frac { 2g^2 Q_c \big(14 \Omega^2 - 8\omega \Omega + 5\omega^2 \big) } {27{\omega}^{7}R}}\cos(\phi)
	+ \frac{2 g^2 \big(38\Omega^2 - 8\omega\Omega + 107\omega^2\big)}{3375 \omega^5}
	\end{aligned},
	\\[0.6ex]
	&\begin{aligned}
	\bar{\Phi}_3 = {\frac {c g \Lambda \big(128\Omega^2 + 50\omega\Omega -45 \omega^2\big)}{288 {\omega}^{4} R}} \sin(\phi)
	+ {\frac {23{g}^{3}\Lambda \big( 646 \Omega - 283 \omega \big)}{18000 {\omega}^{6} R}} \cos ( \phi)
	\end{aligned},
	\\[0.6ex]
	&\begin{aligned}
	\bar{\Phi}_4 = & - {\frac {5 cg^2 Q_c }{36 {\omega}^{6} R}} \sin( \phi )
	-{\frac {\big(428 g^4 + 6075 \omega^6 \Omega^2 \Lambda^2 \big)  Q_c}{5400 {\omega}^{9} R}}
	\cos ( \phi)
	+ {\frac { {g}^{4} }{18 {\omega}^{7}}}\cos ( 2 \phi) 
	\\ &
	+ \frac{\big(3\Omega^2 + 2 \omega^2\big)Q_c^2}{4 \omega^5}
	+ \frac{\big(R^2 + \Lambda^2\big)\omega}{4}
	+ \frac{ \big(3R^2 + 10\Lambda^2\big) + 2c^2\Omega^2}{16 \omega}
	+ \frac{137 g^4}{6750 \omega^7}
	\end{aligned}
	\end{align*}
\end{subequations}

\subsection{In-plane vibrations - primary resonance}
Coefficients in equation for $\dot{R}$ in Eq.\eqref{eq:in-Rphi} ($\mathcal{R}_3$ is omitted for brevity):
\begin{subequations}
	\begin{align*}
	& \begin{aligned}
	\mathcal{R}_1 = - \frac{ (\gamma_{1}\omega^2 \Omega \Lambda + \gamma_{2}\delta_{1} \Omega - \omega^2 \delta_{5} ) (\Omega - 3 \omega)}{4 \omega^4} \cos(\phi)
	- \frac{\gamma_{1} R}{2}
	\end{aligned},
	\\[0.5ex] &
	\begin{aligned}
	\mathcal{R}_2 =& \frac{1}{72 \omega^5}\bigg(
	9  \Big(  ( - \gamma_{{3}}+3  \gamma_{{6}} ) \big( 4 { \delta_{{1}}}^{2}{\Omega}^{2}+ \left( {\Lambda}^{2}+{R}^{2} \right) {\omega}^{4} \big) {\Omega}^{2}  - \gamma_{{3}} {\omega}^{4} \big( {\Lambda}^{2}{\Omega}^{2}+ ( 2 \Omega-\omega) \omega {R}^{2} \big)  \Big) \Lambda
	\bigg.
	\\ & \qquad \bigg.
	+ 9 \left( 8 \gamma_{{5}} \omega \Omega \Lambda
	-\gamma_{{1}}\gamma_{{2}} \right) \delta_{{1}}\omega \Omega
	+ {\gamma_{{2}}}^{2}\omega \Lambda\left( -8 \Omega +11 \omega \right) 
	-9 \gamma_{{1}}{\omega}^{3} \left(\gamma_{{1}}  \Omega \Lambda  -\delta_{{5}} \right) 
	\bigg)\sin(\phi)
	\\& 
	+ \frac{\big( \gamma_{2}(\delta_{3} + \delta_{4}\Omega^2) + \omega^2 (\gamma_{4}\Omega^2 \Lambda - \delta_{6}\Omega)
		\big)}{2 \omega^3}\cos(\phi)
	+ \frac{\big( ( \gamma_{3} (\Omega-\omega)^2 + (2 \gamma_{3}-3\gamma_{6}) \Omega^2 ) \omega^2 \Lambda^2 - \gamma_{2}^2 \big) R }{8\omega^3} \sin(2\phi)
	- \frac{\gamma_{4}\Omega R}{2} .
	\end{aligned}
	\end{align*}
\end{subequations}

Coefficients of in equation for $\dot{\phi}$ in  Eq.\eqref{eq:in-Rphi} ($\Phi_3$ is omitted for brevity):
\begin{subequations}
	\begin{align*}
	& \begin{aligned}
	\Phi_0 = \Omega - \omega
	\end{aligned},
	\\[0.5ex] &
	\begin{aligned}
	\Phi_1 = \frac{\big( \omega^2(\gamma_{1}\Omega\Lambda-\delta_{5}) + \gamma_{2}\delta_{1}\Omega \big)(\Omega - 3\omega) }{4\omega^4 R}
	\sin(\phi)
	\end{aligned},
	\\[0.5ex] &
	\begin{aligned}
	\Phi_2 = & 
	\frac{1}{72 \omega^5 R}\bigg(
	9  \Big( ( - \gamma_{{3}}+3  \gamma_{{6}}) \big( 4 { \delta_{{1}}}^{2}{\Omega}^{2}+ ( {\Lambda}^{2}+ 3 {R}^{2} ) {\omega}^{4} \big) {\Omega}^{2}  - \gamma_{{3}} {\omega}^{4} \big( {\Lambda}^{2}{\Omega}^{2}- ( 2 \Omega - 5\omega) \omega {R}^{2} \big)  \Big) \Lambda
	\bigg.
	\\ & \qquad \bigg.
	+ 9 \left( 8 \gamma_{{5}} \omega \Omega \Lambda
	-\gamma_{{1}}\gamma_{{2}} \right) \delta_{{1}}\omega \Omega
	+ {\gamma_{{2}}}^{2}\omega \Lambda\left( -8 \Omega +11 \omega \right) 
	-9 \gamma_{{1}}{\omega}^{3} \left(\gamma_{{1}}  \Omega \Lambda  -\delta_{{5}} \right) 
	\bigg)\cos(\phi) 
	\\ &
	- \frac{\big(\gamma_{2}(\delta_{3} + \delta_{4}\Omega^2) +\omega^2\Omega(\gamma_{4}\Omega\Lambda-\delta_{6})\big)} {2 \omega^3 R} \sin(\phi)
	+ \frac{\big( ( \gamma_{3} (\Omega-\omega)^2 + (2 \gamma_{3}-3\gamma_{6}) \Omega^2 ) \omega^2 \Lambda^2 - \gamma_{2}^2 \big) }{8\omega^3} \cos(2\phi)
	\\ &
	+ \frac{\gamma_{1}^2}{8 \omega}
	+ \frac{\gamma_{2}^2(2\Omega + \omega)}{36 \omega^4}
	+ \gamma_{3}\bigg(
	\frac{\omega(\Lambda^2 + R^2)}{4} 
	+ \frac{(2\delta_{1}^2 + \omega^2 \Lambda^2)\Omega^2}{4 \omega^3}
	\bigg)
	- \gamma_{6} \bigg(
	\frac{3(2\Lambda^2 + R^2) \Omega^2}{8 \omega}
	+ \frac{3 \delta_{1}^2 \Omega^4}{2 \omega^5}
	\bigg)
	- \frac{\gamma_{5} \delta_{1} \Omega^2}{\omega^3}
	.
	\end{aligned}
	\end{align*}
\end{subequations}


\bibliographystyle{elsarticle-num} 
\bibliography{biblio} 


\end{document}